\documentclass[aps,prd,showpacs,twocolumn,floatfix,nofootinbib,superscriptaddress,
amsmath,amssymb,amsfonts,]{revtex4-1} 

\usepackage{bm}
\usepackage{slashed}
\usepackage{graphicx}

\usepackage{amsmath,amssymb,amsfonts,graphicx}
\usepackage{dsfont, mathrsfs}
\usepackage{bm}
\usepackage{slashed}
\usepackage{subfigure}
\usepackage[svgnames]{xcolor}
\usepackage{placeins}
\usepackage{array}
\usepackage{natbib}
\usepackage{url}
\usepackage[utf8]{inputenc}

\makeatletter
\newcommand{\thickhline}{%
    \noalign {\ifnum 0=`}\fi \hrule height 1pt
    \futurelet \reserved@a \@xhline
}
\newcolumntype{"}{@{\hskip\tabcolsep\vrule width 1pt\hskip\tabcolsep}}
\makeatother

\usepackage{placeins}

\usepackage{xspace}

\newcolumntype{L}[1]{>{\raggedright\let\newline\\\arraybackslash\hspace{0pt}}m{#1}}
\newcolumntype{C}[1]{>{\centering\let\newline\\\arraybackslash\hspace{0pt}}m{#1}}
\newcolumntype{R}[1]{>{\raggedleft\let\newline\\\arraybackslash\hspace{0pt}}m{#1}}
\newcommand{\Eq}[1]{Eq.~(\ref{#1})}

\newcommand{\al}[1]{\begin{align} #1 \end{align}}
\newcommand{\non}{\nonumber}
\newcommand{\vect}[1]{\boldsymbol{#1}}

\def\SA{\sphericalangle}
\def\fh{\varphi_h}

\def\fs{\varphi_s}

\def\fS{\varphi_S}
\def\fR{\varphi_R}

\def\fK{\varphi_{k}}

\def \fk{\varphi_{k}}

\def\fRK{\varphi_{RK}}
\def\kT{\vect{k}_T}

\def\RT{\vect{R}_T}

\def\psq{p_\perp^2}
\newcommand{\psqn}[1]{p_{#1\perp}^2}

\def\PP{\vect{P}_\perp}
\newcommand{\PPn}[1]{\vect{P}_{#1\perp}}

\newcommand{\pe}[1]{\vect{p}_{{#1}\perp}}
\newcommand{\Pe}[1]{\vect{P}_{{#1}\perp}}

\def\sT{\vect{s}_T}

\newcommand{\sq}[1]{\vect{s}_{#1}}

\newcommand{\ImL}{0.85\columnwidth}
\newcommand{\GapCapt}{\vspace{-5pt}}
\newcommand{\GapSubf}{\vspace{-8pt}}
\begin{document}

\title{Dihadron fragmentation functions in the quark-jet model:\\ Transversely polarized quarks}

\preprint{ADP-17-32/T1038}

\author{Hrayr~H.~Matevosyan}
\thanks{ORCID: http://orcid.org/0000-0002-4074-7411}
\affiliation{ARC Centre of Excellence for Particle Physics at the Terascale,\\ 
and CSSM, Department of Physics, \\
The University of Adelaide, Adelaide SA 5005, Australia
\\ http://www.physics.adelaide.edu.au/cssm
}

\author{Aram~Kotzinian}
\thanks{ORCID: http://orcid.org/0000-0001-8326-3284}
\affiliation{Yerevan Physics Institute,
2 Alikhanyan Brothers St.,
375036 Yerevan, Armenia
}
\affiliation{INFN, Sezione di Torino, 10125 Torino, Italy
}

\author{Anthony~W.~Thomas}
\thanks{ORCID: http://orcid.org/0000-0003-0026-499X}
\affiliation{ARC Centre of Excellence for Particle Physics at the Terascale,\\     
and CSSM, Department of Physics, \\
The University of Adelaide, Adelaide SA 5005, Australia
\\ http://www.physics.adelaide.edu.au/cssm
}

\begin{abstract}
Within the most recent extension of the  quark-jet hadronization framework, we explore the transverse-polarization-dependent dihadron fragmentation functions (DiFFs) $H_1^\SA$ and $H_1^\perp$ of a quark into $\pi^+\pi^-$ pairs.  Monte Carlo (MC) simulations are employed  to model polarized quark hadronization and calculate the corresponding number densities. These, in turn, are used to extract the Fourier cosine moments of the DiFFs $H_1^\SA$ and $H_1^\perp$.  A notable finding is that there are previously unnoticed apparent discrepancies between the definitions of the so-called interference DiFF (IFF) $H_1^\SA$, entering the cross sections for two-hadron semi-inclusive electroproduction, and those involved in the production of two pairs of hadrons from back-to-back jets in  electron-positron annihilation. This manuscript completes the studies of all four leading twist DiFFs for unpolarized hadron pairs within the quark-jet framework, following our previous work on the helicity-dependent DiFF $G_1^\perp$.
\end{abstract}



\date{\today}                                           

\maketitle

\section{Introduction}
\label{SEC_INTRO}

The study of the transverse-polarization-dependent dihadron fragmentation functions (DiFFs)~\cite{Bianconi:1999cd, Bianconi:1999uc}, and of the interference DiFF (IFF) in particular, has gained a lot of attention in recent years, both in theory and in experiment. The chief reason is that the IFF is a crucial ingredient in accessing the so-called transversity parton distribution function (PDF) in dihadron production in semi-inclusive deep inelastic scattering (SIDIS)~\cite{Radici:2001na,Bacchetta:2003vn,Bacchetta:2011ip, Bacchetta:2012ty, Pisano:2015wnq}. This is important, since the transversity PDF, which at leading approximation describes the correlations of the transverse polarizations of the nucleon and its constituent partons, is the least well-known leading-twist collinear PDF due to its chiral-odd nature. The recent phenomenological analysis in Refs.~\cite{Courtoy:2012ry,Radici:2015mwa} used BELLE measurements~\cite{Vossen:2011fk} of the azimuthal asymmetry involving IFF $H_1^\SA$ in inclusive dihadron pair production in back-to-back jets emanating from $e^+e^-$ annihilation, along with input from Monte Carlo (MC) simulations for the unpolarized DiFF, to fit both the unpolarized DiFF $D_1$ and the IFF $H_1^\SA$. These fits were then used to extract the transversity PDFs from the SIDIS measurements at HERMES~\cite{Airapetian:2008sk} and COMPASS~\cite{Adolph:2012nw,Adolph:2014fjw}. The recent BELLE measurements in Ref.~\cite{Seidl:2017qhp} would provide the information on the unpolarized DiFFs, eliminating the need for the MC input. Nevertheless, systematic improvements of the extraction procedure might benefit from the information that can be gained from the model calculations of the DiFFs, for example by providing some guidance concerning the parametrizations of the unpolarized DiFFs and IFFs.

 The quark-jet framework is based on the original model of the collinear hadronization of an unpolarized quark~\cite{Field:1976ve,Field:1977fa}. Over the past several years it has been developed to include the production of kaons and other hadrons~\cite{Ito:2009zc,Matevosyan:2010hh,Matevosyan:2011ey,PhysRevD.86.059904}, as well as the transverse momentum dependence~\cite{Matevosyan:2011vj} by utilizing MC simulations of the hadronization process and the Nambu--Jona-Lasinio (NJL) quark model~\cite{Nambu:1961tp,Nambu:1961fr} for calculating the input elementary fragmentation functions. We also studied DiFFs~\cite{Matevosyan:2013aka, Matevosyan:2013nla, Matevosyan:2013eia} within the framework, with a simplistic treatment of the quark polarization to access the so-called single-hadron Collins fragmentation function (FF) and the IFF. A self-consistent treatment of the quark polarization within the quark-jet picture was recently developed in Ref.~\cite{Bentz:2016rav} and used to calculate the Collins fragmentation functions of a polarized light quark into pions~\cite{Matevosyan:2016fwi}.
 
Following our previous work on calculations of the helicity-dependent DiFF in~\cite{Matevosyan:2017alv} within the quark-jet framework, here we extend the study to the DiFFs involving the correlations between the transverse polarization of the fragmenting quark and the transverse momenta of the two hadrons. The goal is to provide a description of the complete set of the leading-twist two-hadron fragmentation functions for unpolarized hadron pairs within a self-consistent framework of polarized quark hadronization and to complement the description of the single hadron fragmentation functions in Ref.~\cite{Matevosyan:2016fwi} within the same approach.

 This paper is organized in the following way.  In Sec.~\ref{SEC_DIFF_FORM} we summarize the DiFF kinematics and formal definitions. In Sec.~\ref{SEC_MC} we derive the expressions for extracting the moments of DiFFs from the polarized number density. The numerical results for the DiFFs  are presented in Sec.~\ref{SEC_NJL_JET}, while their validation tests against explicit integral expressions for the two-hadron emission process are detailed in Sec.~\ref{SEC_VALIDATION}. The final conclusions are presented in Sec.~\ref{SEC_CONCLUSIONS}.

\section{Formalism}
\label{SEC_DIFF_FORM}
 
\subsection{Kinematics and definitions}
\label{SUBSEC_DIFF_DEF}
  We begin the discussion by reviewing the kinematics and field theoretical definitions of the DiFFs. A more detailed description of the kinematics and the formalism can be found, e.g. in Ref.~\cite{Matevosyan:2017alv}. Here we consider the fragmentation of a quark $q$ with momentum $k$ and mass $m$ into a pair of hadrons $h_1,h_2$ with masses $M_1,M_2$ and corresponding momenta $P_1$ and $P_2$.  The total and the relative transverse momenta are defined as
\al
{
&
 P \equiv P_h = P_1 + P_2,
 \\
 &
 R = \frac{1}{2}( P_1 - P_2).
 }

 The $\hat{z}$ directions for the two relevant coordinate systems labeled $\perp$  and $T$ are defined using either the 3-momentum of the quark $\vect{k}$ or the total 3-momentum of the  hadron pair $\vect{P}$, respectively. The components of 3-vectors perpendicular to the $\hat{z}$ directions in the two systems are denoted with subscripts $\perp$ and $T$. The total and the relative light-cone momentum\footnote{ We define the light-cone components of a 4-vector $a$ as $a^\pm = \frac{1}{\sqrt{2}}(a^0 \pm a^3)$.} fractions are expressed in terms of those for individual hadrons $z_i = {P_i^+}/{k^+}$ as
\al
{
 z &= z_1  + z_2 ,
\\
 \xi &= \frac{z_1}{z}  = 1- \frac{z_2}{z}.
}

  The transverse components of the relevant momenta in the two systems are related using the following expressions
\al
{
 \vect{P}_{1T} &= \PPn{1}  + z_1 \kT,
 \\
 \vect{P}_{2T} &= \PPn{2}  + z_2 \kT,
 \\
 \kT &= -\frac{\PP}{z},
 \\
 \RT &= \frac{z_2 \PPn{1} - z_1 \PPn{2}}{z} = (1-\xi) \PPn{1} - \xi \PPn{2}.
}

 The magnitude of $\RT$ is often replaced by the invariant mass $M_h$ of the hadron pair 
\al
{
 M_h^2 &= P_h^2,
 \\
R_T^2 & =  \xi (1-\xi) M_h^2 - M_1^2 (1-\xi) - M_2^2 \xi.
}

The field-theoretical definitions of the DiFFs~\cite{Bianconi:1999cd, Radici:2001na, Boer:2003ya} are given using projections of the quark-quark correlator $\Delta$ with various  Dirac operators
\al
{
\label{EQ_DELTA_UNP}
 \Delta^{[\gamma^+]} =& D_1(z, \xi, \kT^2, \RT^2, \kT \cdot \RT),
\\ \label{EQ_DELTA_LIN}
  \Delta^{[\gamma^+\gamma_5]} 
  =& \frac{\epsilon_T^{ij} R_{Ti} k_{Tj}}{M_1 M_2} G_1^\perp(z, \xi, \kT^2, \RT^2, \kT \cdot \RT),
 \\ \label{EQ_DELTA_TRANSV}
 \Delta^{[i \sigma^{i+} \gamma_5]} 
  =& \frac{\epsilon_T^{ij} R_{Tj}}{M_1 + M_2} H_1^\SA(z, \xi, \kT^2, \RT^2, \kT \cdot \RT) 
  \\ \non
& + \frac{\epsilon_T^{ij}  k_{Tj}}{M_1 + M_2} H_1^\perp(z, \xi, \kT^2, \RT^2, \kT \cdot \RT),
}
where $D_1$ is the unpolarized, $G_1^\perp$ is the helicity-dependent, $H_1^\SA$ is the IFF, and  $H_1^\perp$ is the analogue of the single-hadron Collins fragmentation function. The tensor $\epsilon_T^{ij}\equiv \epsilon^{-+ij}$ is the "transverse" Levi-Civit\`a tensor, and we use the convention $\epsilon_T^{12}=+1$.

\subsection{Fourier moments of DiFFs}
\label{SUBSEC_DIFF_FOURIER}
The SIDIS cross section in general can be decomposed in terms of the convolutions of various PDFs with an infinite series of Fourier moments of the fully unintegrated DiFFs; see Ref.~\cite{Gliske:2014wba}. It is clear that the angular dependence of the DiFFs defined in Eqs.~(\ref{EQ_DELTA_UNP})-(\ref{EQ_DELTA_TRANSV}) is encoded in the argument $\kT \cdot \RT \propto \cos(\fRK)$, where $\fRK \equiv \fR-\fK$ is the difference between the azimuthal angles $\fR$  and $\fK$  of the vectors $\RT$ and $\kT$. Thus, the Fourier decomposition of these DiFFs only involves their cosine moments. We define the nth cosine moments for different integrated DiFFs as
\al
{
 \label{EQ_D1_MOM_MH}
 D_1^{[n]}(z, M_h^2)
  = & \int  d \xi  \int d \fR  \int d^2 \kT  
\\ \non
 &\times \cos(n \cdot \fRK) \ D_1(z, \xi, \kT^2, \RT^2, \kT \cdot \RT),
}
\al
{
 \label{EQ_G1_MOM_MH}
 G_1^{\perp,[n]}(&z, M_h^2)
  =  \int d \xi  \int d \fR  \int d^2 \kT  
\\ \non
 & \times \cos(n \cdot \fRK)\ k_T R_T  \ G_1^\perp(z, \xi, \kT^2, \RT^2, \kT \cdot \RT).
}
\al
{
 \label{EQ_HANG_MOM_MH}
H_1^{\SA, [n]}(z,& M_h^2)
  =  \int  d \xi  \int d \fR  \int d^2 \kT  
\\ \non
 &\times \cos(n \cdot \fRK) \ R_T\ H_1^\SA(z, \xi, \kT^2, \RT^2, \kT \cdot \RT),
}
\al
{
 \label{EQ_HPERP_MOM_MH}
 H_1^{\perp, [n]}(z,& M_h^2)
  =  \int  d \xi  \int d \fR  \int d^2 \kT  
\\ \non
 &\times \cos(n \cdot \fRK) \ k_T\ H_1^\perp(z, \xi, \kT^2, \RT^2, \kT \cdot \RT).
}

 The integrated cross section for two back-to-back dihadron pairs produced in $e^+e^-$ annihilation involves only single  Fourier cosine moments~\cite{Boer:2003ya}:
\al
{
 \label{EQ_D1_EE}
    D_1(z, M_h^2) & \equiv D_1^{[0]}(z, M_h^2),
}
\al
{
 \label{EQ_GP_EE}
    G_1^{\perp}(z, M_h^2) & \equiv  G_1^{\perp,[1]}(z, M_h^2),
}
\al
{
 \label{EQ_HANG_EE}
    H_1^{\SA, e^+e^-}(z, M_h^2) & \equiv H_1^{\SA, [0]}(z, M_h^2).
}
%

 On the other hand, the situation is different for the $H_1^{\SA}$ appearing in the integrated SIDIS cross section, used in extracting the transversity PDF using the "collinear framework" in Refs.~\cite{Radici:2001na,Radici:2015mwa}. In the original derivation of the corresponding expression~\cite{Radici:2001na}, only the zeroth cosine moment enters the definition of the integrated IFF. This omits a contribution to the cross section involving $H_1^\perp$, as can be easily seen by following the derivation steps of the expression in Eq.~(17) in Ref.~\cite{Radici:2001na} starting from the fully unintegrated cross section in Eq.~(10). This was also  shown in Eqs.~(17-20) and (46) in Ref.~\cite{Bacchetta:2003vn}, where the derivation of the SIDIS cross section was performed using a quark-quark correlator that was first integrated over the total transverse momentum of the hadron pair (same as the quark transverse momentum in the $T$ system). The relevant expression for the IFF entering the SIDIS cross section is
\al
{
 \label{EQ_HANG_SIDIS}
    H_1^{\SA, SIDIS}(z, M_h^2)  \equiv H_1^{\SA, [0]}(z, M_h^2) +   H_1^{\perp, [1]}(z, M_h^2) \, .
}

 Analogously, a similar reduction of the unintegrated cross section into a modulation involving the convolution of the transversity PDF and $H_1^{\perp}$ involves
\al
{
 \label{EQ_HPERP_SIDIS}
    H_1^{\perp, SIDIS}(z, M_h^2)  \equiv H_1^{\perp, [0]}(z, M_h^2) +   H_1^{\SA, [1]}(z, M_h^2) \, ,
}
This term can be isolated by integrating the $\sin(\fh+\fS)$ weighted SIDIS cross section expression in  Eq.~(10) in Ref.~\cite{Radici:2001na}, where $\fh$ and $\fS$ are the azimuthal angles of the hadron pair's total transverse momentum and the transverse polarization of the target nucleon. We also note that here too the integrated unpolarized DiFF is just the zeroth moment of the unintegrated one, similar to the $e^+e^-$ case.
 
 Thus, it appears that there could be a difference between the IFFs extracted from $e^+e^-$ and those entering the SIDIS cross section, because of the admixture of the first Fourier cosine moment of $H_1^\perp$ in the latter. It is clearly important to understand the magnitude of this effect, at least using some model calculations, as they can potentially impact the phenomenological extraction of the transversity PDF from data. We have to emphasize here, that there may be missing terms in the expression for the unintegrated cross section of $e^+e^-$ annihilation in Ref.~\cite{Boer:2003ya}, that might contribute to the integrated cross section and resolve the  discrepancies between the two definitions of IFFs in Eqs.~(\ref{EQ_HANG_EE}) and (\ref{EQ_HANG_SIDIS}). Nonetheless, the results for the corresponding cosine moments of the integrated DiFFs presented in this paper will remain true.
  
\section{Extracting DiFFs from Monte Carlo simulations}
\label{SEC_MC}

 We use MC simulations of the polarized quark hadronization process, as described in detail in Refs.~\cite{Matevosyan:2011ey,Matevosyan:2011vj, Matevosyan:2013aka, Matevosyan:2013nla, Matevosyan:2013eia, Matevosyan:2016fwi, Matevosyan:2017alv}. We calculate various number densities by averaging over a large number of MC simulation events of quark hadronization. The DiFFs are extracted from these numbers densities using the corresponding modulations with respect to azimuthal angles. The expression for the number density for the production of two unpolarized hadrons in polarized quark fragmentation can be encoded using the definitions in Eqs.~(\ref{EQ_DELTA_UNP})-(\ref{EQ_DELTA_TRANSV}):
\al
{
\label{EQ_F_VEC}
 F(z, \xi , &\kT, \RT; \vect{s}) = D_1(z, \xi, \kT^2, \RT^2, \kT \cdot \RT) 
 \\ \non
 &+ s_L \frac{ (\RT \times  \kT)\cdot \vect{\hat{z}} }{M_1 M_2} G_1^\perp(z, \xi, \kT^2, \RT^2, \kT \cdot \RT)
 \\ \non
 & + \frac{ (\sT \times \RT)\cdot \vect{\hat{z}} }{M_1 + M_2} H_1^\SA(z, \xi, \kT^2, \RT^2, \kT \cdot \RT) 
\\ \non
& + \frac{ (\sT \times \kT)\cdot \vect{\hat{z}} }{M_1 + M_2} H_1^\perp(z, \xi, \kT^2, \RT^2, \kT \cdot \RT) \, .
}
This can also be written in terms of azimuthal angles $\fK$ and $\fR$ of vectors $\kT$ and $\RT$
\al
{
\label{EQ_F_ANG}
F&(z, \xi , \kT, \RT; \vect{s}) = D_1(z, \xi, \kT^2, \RT^2, \cos(\fRK)) 
 \\ \non
 &
 - s_L \frac{ R_T k_T \sin(\fRK) }{M_1 M_2} G_1^\perp(z, \xi, \kT^2, \RT^2, \cos(\fRK))
 \\ \non
 &
 + s_T\frac{  R_T \sin(\fR-\fs)}{M_1 + M_2} H_1^\SA(z, \xi, \kT^2, \RT^2, \cos(\fRK)) 
 \\ \non
 &
 + s_T\frac{  k_T \sin(\fK-\fs) }{M_1 + M_2} H_1^\perp(z, \xi, \kT^2, \RT^2, \cos(\fRK)).
}

The Fourier cosine moment of the DiFFs in Eqs.~(\ref{EQ_D1_MOM_MH})-(\ref{EQ_HPERP_MOM_MH}) then can be obtained by integrating the number density multiplied by specific trigonometric factors
\al
{
D_1^{[n]}(z, M_h^2)=& \int d \xi  \int d \fR  \int d^2 \kT 
\\ \non 
 &\times \cos(n \cdot \fRK) \ F(z, \xi , \kT, \RT; \vect{s}),
}
\al
{
\frac{G_1^{\perp,[n]}(z, M_h^2)}{M_1 M_2} = -&\frac{1}{s_L}  \int d \xi  \int d \fR  \int d^2 \kT 
 \\ \non 
 &\times \ \frac{\cos(n \cdot \fRK) }{\sin(\fRK)} F(z, \xi , \kT, \RT; \vect{s}),
}
\al
{
\label{EQ_HANG_EXTRACT}
&\frac{H_1^{\SA,[n]}(z, M_h^2)}{M_1+ M_2} = \frac{2}{s_T}  \int d \xi  \int d \fR  \int d^2 \kT 
 \\ \non 
 &\quad  \quad \quad \times \ \cos(\fk-\fs)  \frac{\cos(n \cdot \fRK) }{\sin(\fRK)} F(z, \xi , \kT, \RT; \vect{s}),
}
\al
{
\label{EQ_HPERP_EXTRACT}
&\frac{H_1^{\perp,[n]}(z, M_h^2)}{M_1+ M_2} = -\frac{2}{s_T}  \int d \xi  \int d \fR  \int d^2 \kT 
 \\ \non 
 & \quad \quad \quad \times \ \cos(\fR-\fs)  \frac{\cos(n \cdot \fRK) }{\sin(\fRK)} F(z, \xi , \kT, \RT; \vect{s}).
}

Using this method, we can extract the zeroth moments of $H_1^\SA$, $H_1^\perp$, with the former entering the cross section expression for $e^+e^-$ annihilation; see \Eq{EQ_HANG_EE}. We can also extract the first moments and calculate the expressions for $H_1^\SA, H_1^\perp$ entering the SIDIS  cross section using Eqs.~(\ref{EQ_HANG_SIDIS}) and (\ref{EQ_HPERP_SIDIS}). On the other hand, we can also directly extract these two DiFFs from the polarized number densities
\al
{
\label{EQ_HANG_SIDIS_EXTRACT}
 \frac{H_1^{\SA, SIDIS}(z, M_h^2)}{M_1+ M_2}  =& \frac{2}{s_T}  \int d \xi  \int d \fR  \int d^2 \kT 
 \\ \non 
 &\times \ \sin(\fR-\fs)  F(z, \xi , \kT, \RT; \vect{s}),
}
\al
{
\label{EQ_HPERP_SIDIS_EXTRACT}
 \frac{H_1^{\perp, SIDIS}(z, M_h^2)}{M_1+ M_2}  = &\frac{2}{s_T}  \int d \xi  \int d \fR  \int d^2 \kT 
 \\ \non 
 &\times \ \sin(\fk-\fs)  F(z, \xi , \kT, \RT; \vect{s}).
}

We focus on the $M_h^2$ integrated DiFFs in this work, and introduce the following notation for the corresponding dimensionless  functions
\al
{
D_1(z) &\equiv \int d M_h^2 \ D_1(z, M_h^2) ,
\\
\tilde{G}_1^\perp(z) &\equiv  \frac{1}{M_1 M_2} \int d M_h^2 \ G_1^\perp(z, M_h^2)) ,
 \\
\tilde{H}_1^\SA(z) &\equiv  \frac{1}{M_1 + M_2} \int d M_h^2 \ H_1^\SA(z, M_h^2) ,
\\
\tilde{H}_1^\perp(z) &\equiv  \frac{1}{M_1+  M_2} \int d M_h^2 \ H_1^\perp(z, M_h^2) .
}
%

\section{The quark-jet model results}
\label{SEC_NJL_JET}

\begin{figure}[b]
\centering 
\includegraphics[width=0.8\columnwidth]{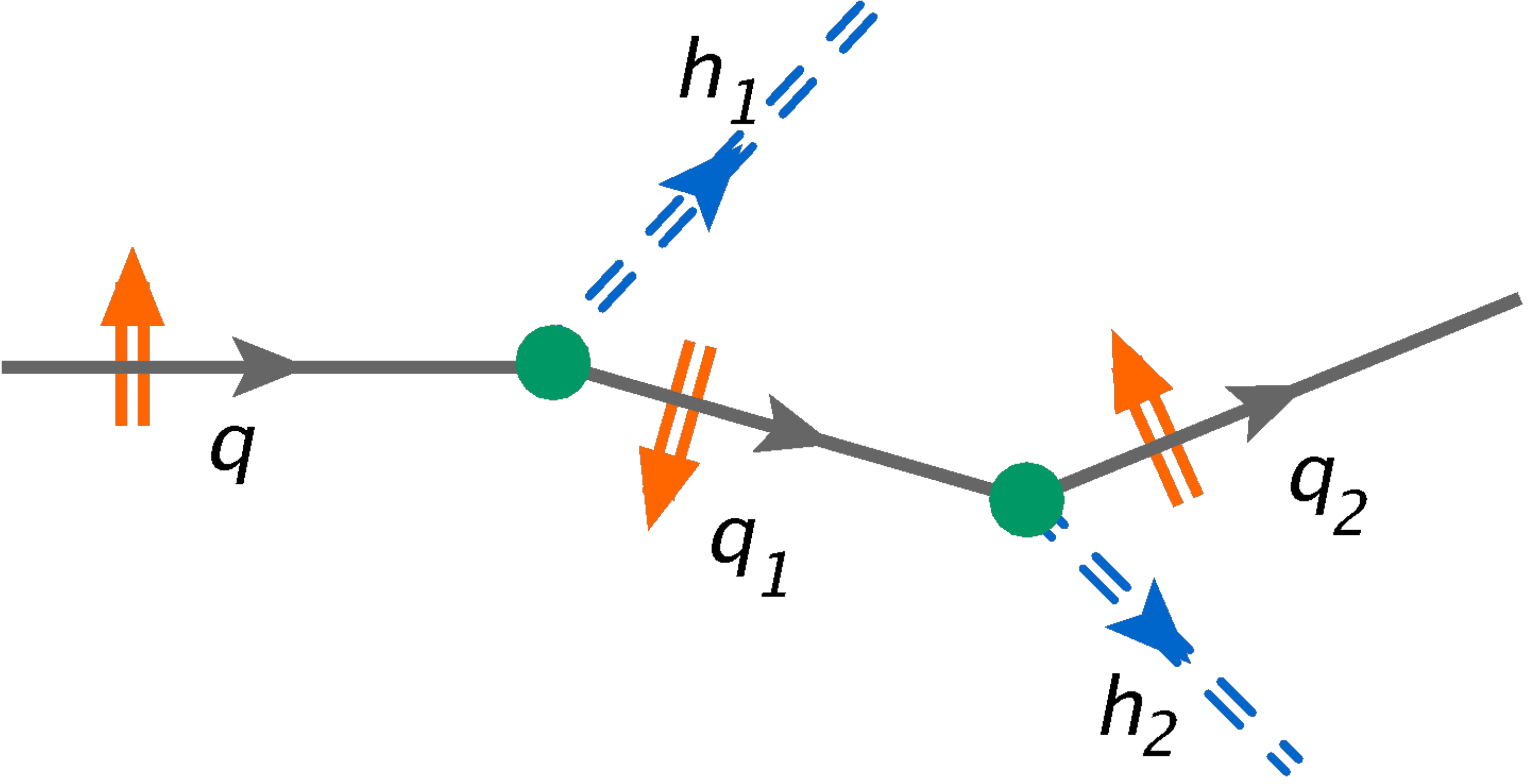}
\GapCapt
\caption{Hadronization within the quark-jet framework.}
\label{PLOT_QUARK_JET}
\end{figure}

 We model the hadronization of a transversely polarized quark within the most recent extension of the quark-jet framework~\cite{Bentz:2016rav, Matevosyan:2016fwi}, in the same manner as for the longitudinally polarized quark in Ref.~\cite{Matevosyan:2017alv}. The emission of hadrons $h_1$, $h_2$  and so forth by a quark within the model is shown schematically in Fig~\ref{PLOT_QUARK_JET}.
The transverse momentum and the light-cone momentum fraction of the remnant quark after each emission are determined using momentum conservation along those directions, while the polarization is determined by its momentum and the polarization of the emitting quark using the spin density matrix formalism. MC simulations of the hadronization process are used to calculate the quark-polarization-dependent number densities as averages over the corresponding numbers of hadron pairs of given momenta. Each hadronization simulation is terminated after a fixed number of emissions, $N_L$.  The remaining input into the quark-jet model for the numerical computations is the eight quark-to-quark and two quark-to-hadron splitting functions (SF) that are needed to sample the types of the produced hadrons, their momenta and the polarization of the remnant quark. In this work we use SFs calculated using the low-energy effective quark model of Nambu--Jona-Lasinio (NJL)~\cite{Nambu:1961tp,Nambu:1961fr}. Specifically, we employ the SFs modified by the $(1-z)^4$ ansatz presented in Ref.~\cite{Matevosyan:2016fwi}, which mimics the effects of the QCD evolution on the unpolarized FFs and DiFFs by shifting the functions towards the lower-$z$ region. Though the QCD evolution equations of the DiFFs~\cite{Ceccopieri:2007ip}, needed in order to compare the results of our low-energy model with experiment, are well known~\cite{Ceccopieri:2007ip} and we applied them to the quark-jet modeled unpolarized DiFFs~\cite{Matevosyan:2013aka}, here we omit  that step for simplicity. Also, we use only one set of input SFs to outline the overall qualitative features of our results, which are mostly independent of these inputs. A detailed comparison of the dependence of the results for the single-hadron FFs on the choice of both the initial model SFs and the $(1-z)^4$ ansatz was presented in Ref.~\cite{Matevosyan:2016fwi}, demonstrating this point. Further, this model only considers the up and down quarks and pions for simplicity, with exact isospin symmetry ($M_u = M_d$ and $m_{\pi^\pm} = m_{\pi^0}$). The flavor of  the initial quark is always set to $u$.

 The polarization of the initial quark in the MC simulations is set to be completely in the transverse direction ($s_L=0$ and $s_T{}=1$) to maximize the signal for extracting $H_1^\SA$ and $H_1^\perp$ that is proportional to $s_T$. It is possible to simultaneously extract the $G_1^\perp$ DiFF by choosing a nonzero value for $s_L$, but this would require higher statistics to achieve the same precision as $s_L^2 + s_T^2 \leq 1$. The results in this section are obtained using $10^{12}$ MC simulations. We choose $100$ discretization points for $z$ in the region $[0,1]$ and $200$ discretization points for the azimuthal angles in region $[0,2\pi)$.
 
\begin{figure}[tb]
\centering 
\includegraphics[width=\ImL]{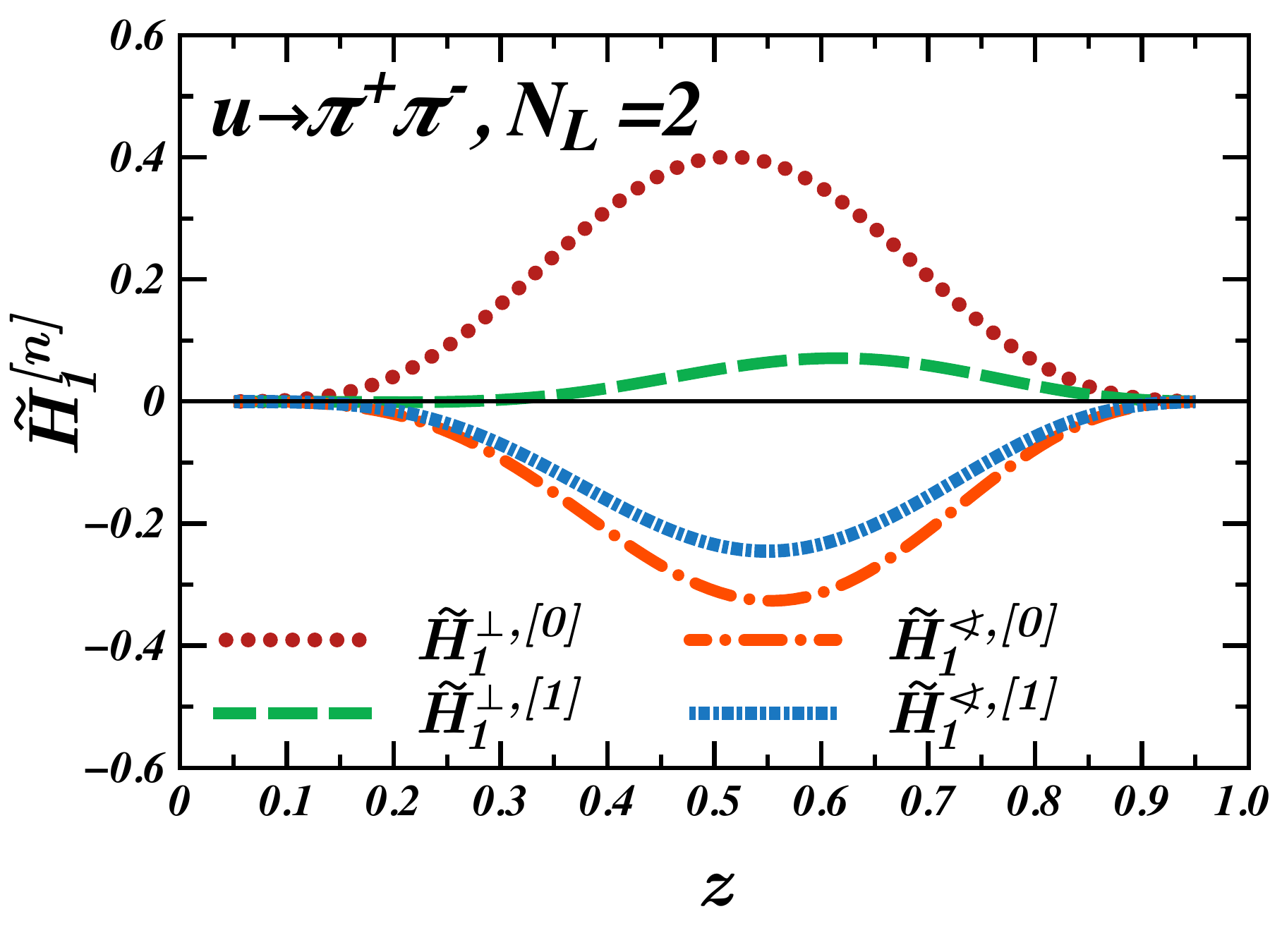}
\GapCapt
\caption{The MC results for the $z$ dependencies of the first two moments of $\tilde{H}_1^\SA(z)$ and $\tilde{H}_1^\perp(z)$ for $\pi^+\pi^-$ pairs, where  $N_L=2$.}
\label{PLOT_H_MOMS_NL2}
\end{figure}

 The first results presented here are for the zeroth and the first moments of $\tilde{H}_1^\SA$ and $\tilde{H}_1^\perp$, as shown in Fig.~(\ref{PLOT_H_MOMS_NL2}). We allow only the minimal possible number of produced hadrons, $N_L=2$, in these simulations. The opposite signs for $\tilde{H}_1^\SA$ and $\tilde{H}_1^\perp$ can be intuitively deduced by considering the orientations of the relative and the total transverse momenta of a back-to-back hadron pair. An interesting observation here is that the relative size of the first moment of $\tilde{H}_1^\SA$ is comparable with sizes of the zeroth moments for both $\tilde{H}_1^\SA$ and $\tilde{H}_1^\perp$. On the other hand, the first moment of $\tilde{H}_1^\perp$ is much smaller than the others in the plot. The importance of this result will become apparent a little later. 

\begin{figure}[!tb]
\centering 
\subfigure[]
 {
\includegraphics[width=\ImL]{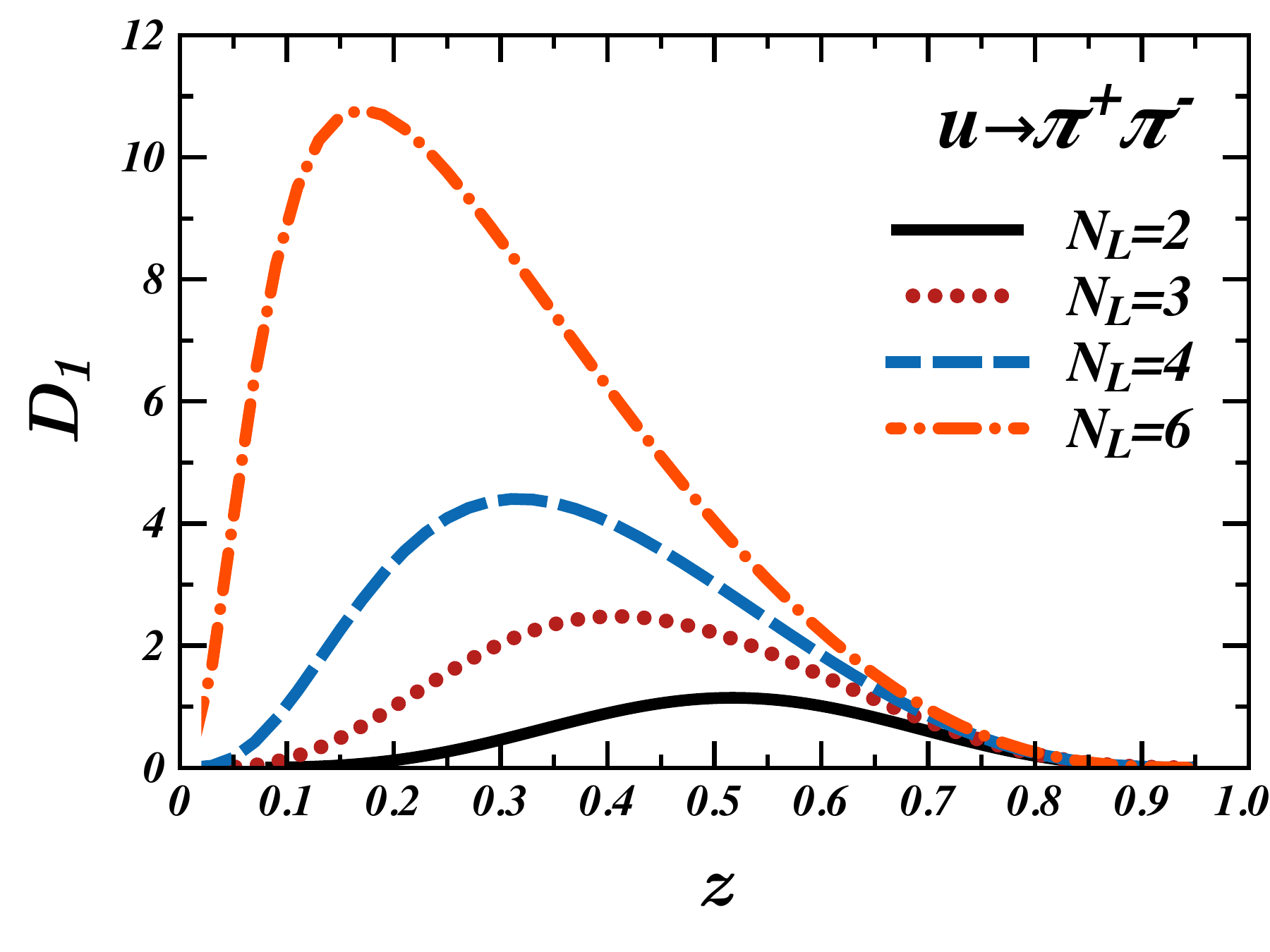}
}
\\ \GapSubf
\subfigure[]
 {
\includegraphics[width=\ImL]{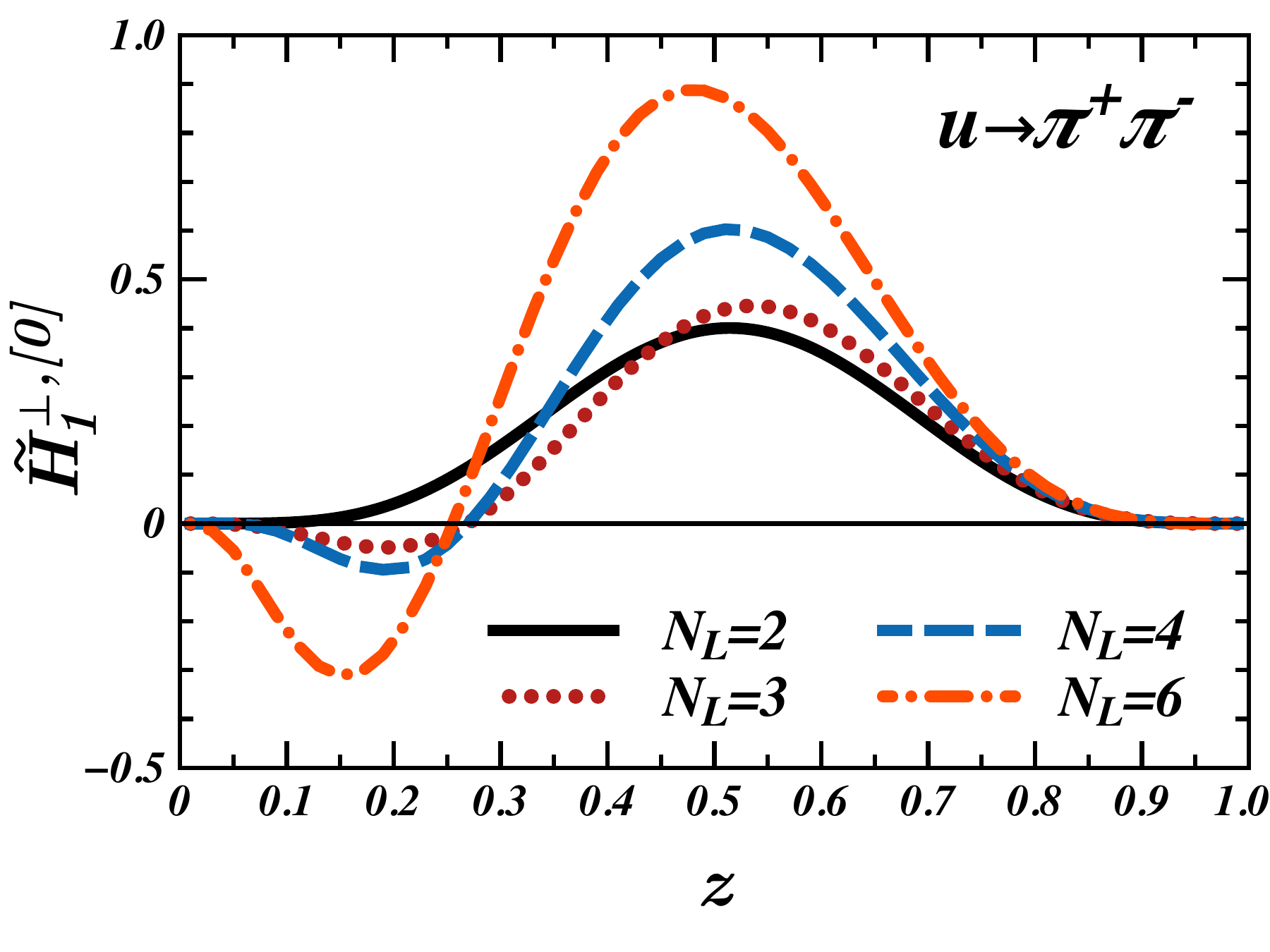}
}
\\ \GapSubf
\subfigure[]
 {
\includegraphics[width=\ImL]{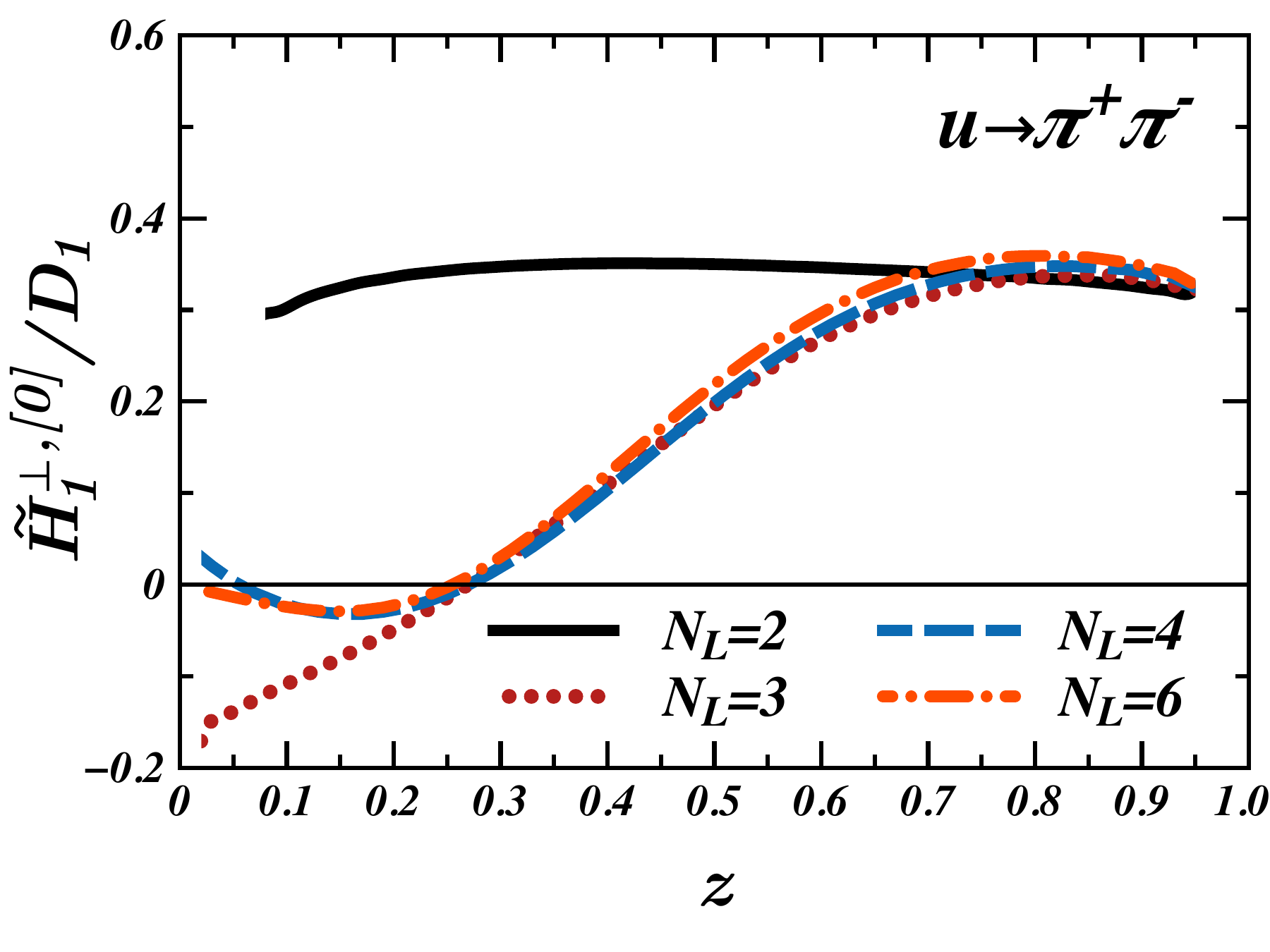}
}
\GapCapt
\caption{The variation of MC results for (a) $D_1(z)$, (b) $\tilde{H}_1^{\perp,[0]}(z)$, and (c) their ratios for $\pi^+\pi^-$ pairs with the increasing values of $N_L$.}
\label{PLOT_D1_NLX}
\end{figure}

\begin{figure}[tb]
\centering 
\subfigure[]
 {
\includegraphics[width=\ImL]{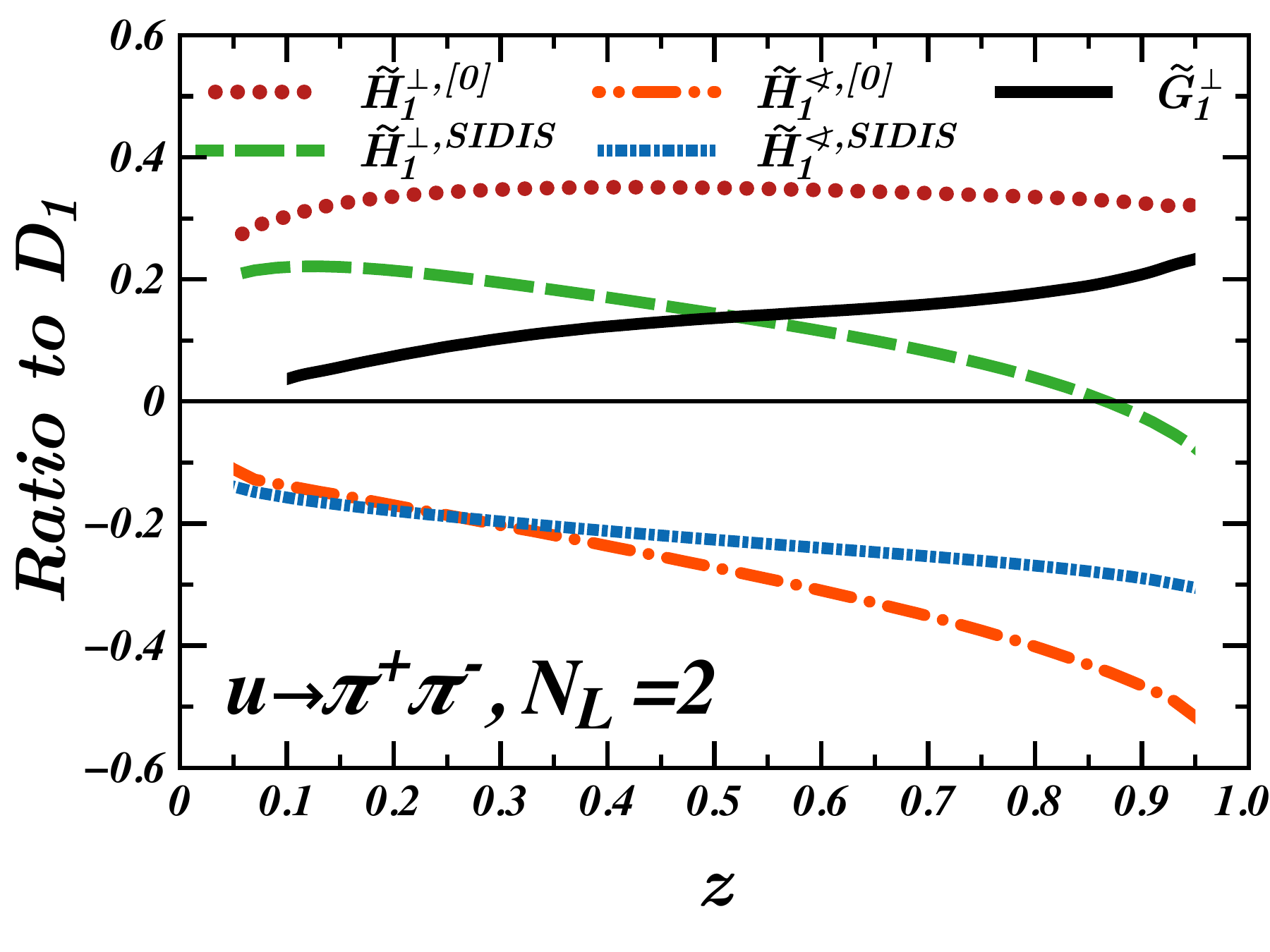}
}
\\ \GapSubf
\subfigure[]
 {
\includegraphics[width=\ImL]{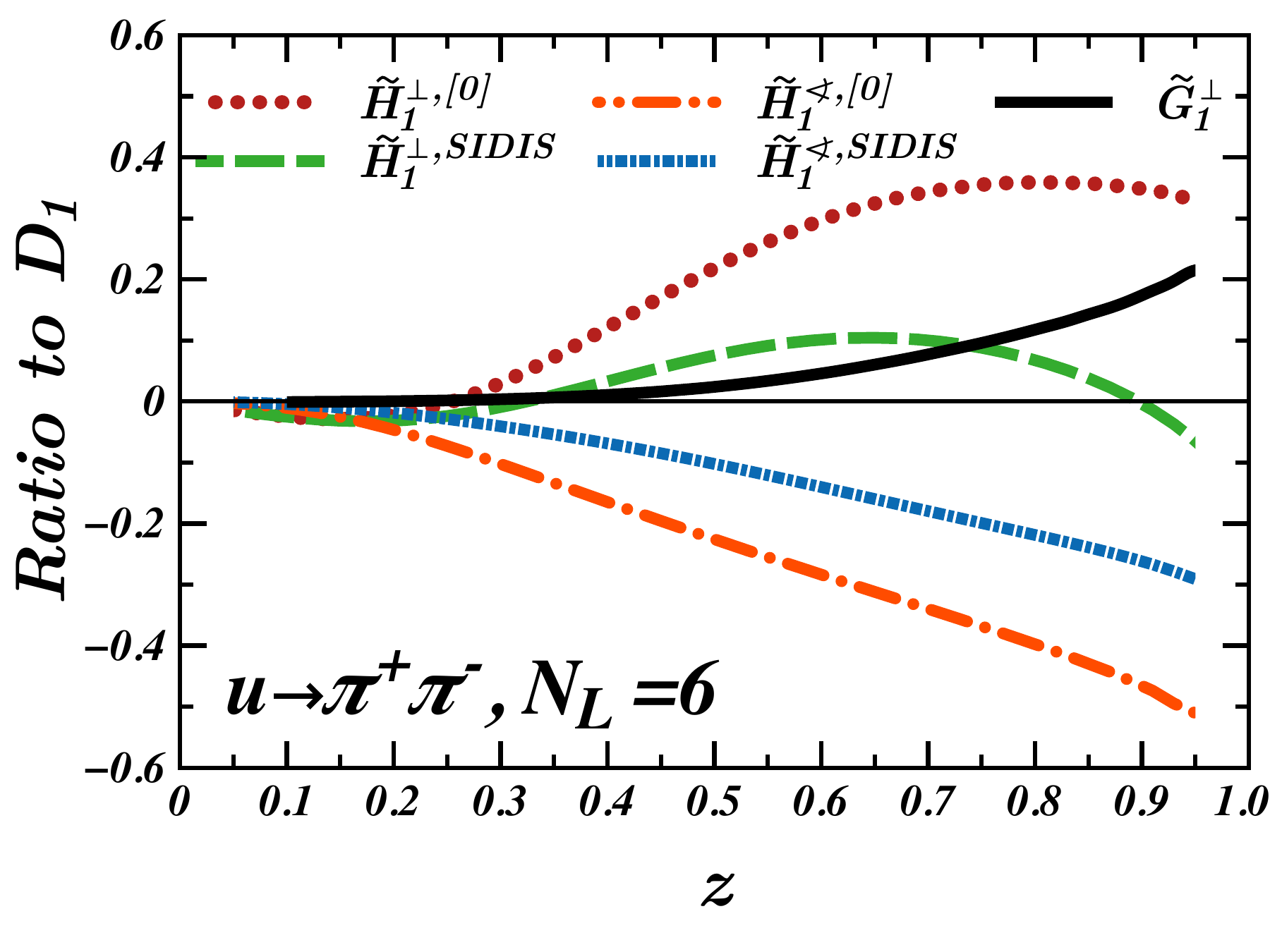}
}
\\ \GapSubf
\subfigure[]
 {
\includegraphics[width=\ImL]{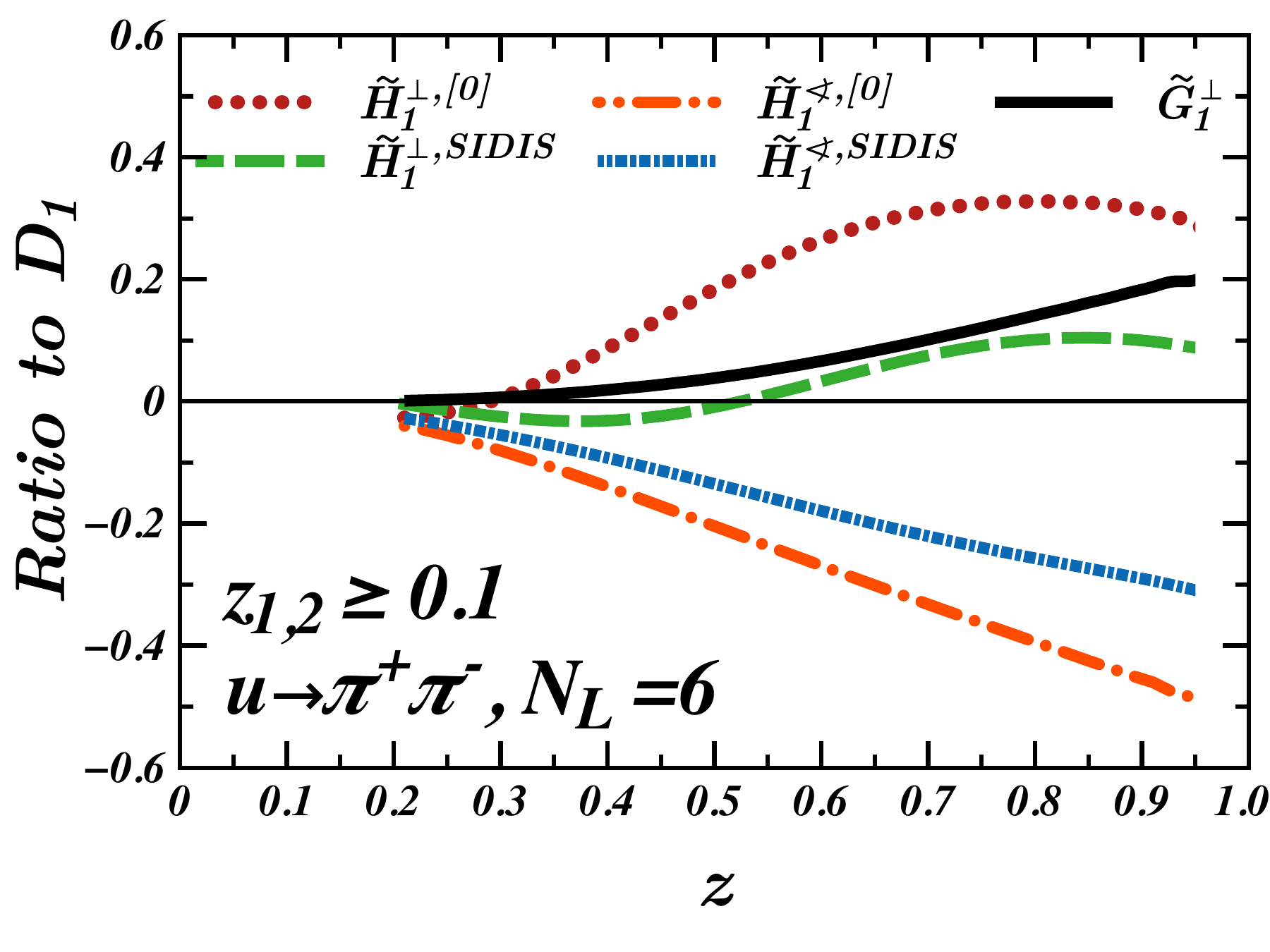}
}
\GapCapt
\caption{The ratios of $\tilde{H}_1^{\SA ,[0]}(z)$, $\tilde{H}_1^{\SA, SIDIS}(z)$, $\tilde{H}_1^{\perp ,[0]}(z)$, $\tilde{H}_1^{\perp, SIDIS}(z)$ and $G^{\perp}(z)$ to $D_1(z)$ for (a) $N_L=2$,  (b) $N_L=6$, and (c) $N_L=6$ with an additional cut $z_{1,2}\geq 0.1$,  for $\pi^+\pi^-$ pairs.}
\label{PLOT_H_RAT_PAIRS}
\end{figure}

 It is important to  present once again the results for the unpolarized DiFF $D_1(z)$ in Fig.~\ref{PLOT_D1_NLX}(a), even though we have already discussed them in the context of extracting the helicity-dependent DiFF $G_1^\perp$ in Ref.~\cite{Matevosyan:2017alv}. We ensure the extracted values of  $D_1$ are the same for different MC simulations as a consistency check. Also, they are important in fully assessing the analyzing powers for the DiFFs shown in Figs.~\ref{PLOT_D1_NLX}(c) and \ref{PLOT_H_RAT_PAIRS}. The $N_L$ dependencies of $\tilde{H}_1^{\perp,[0]}(z)$ and its analyzing power are depicted in Figs.~\ref{PLOT_D1_NLX}(b) and~\ref{PLOT_D1_NLX}(c), respectively. It is clear that the analyzing power saturates rapidly with the increasing $N_L$. The same holds for the analyzing powers of $\tilde{H}_1^{\perp,[1]}(z)$, $\tilde{H}_1^{\SA,[0]}(z)$, and  $\tilde{H}_1^{\SA,[1]}(z)$ not shown here.
 
The plots in Fig.~\ref{PLOT_H_RAT_PAIRS} depict the analyzing powers of $\tilde{H}_1^\SA$ and $\tilde{H}_1^\perp$ for $N_L=2$, $N_L=6$, and $N_L=6$ with additional cuts $z_{1,2}\geq 0.1$ imposed on the hadrons in the pair. The results for both of the moments entering the $e^+e^-$ cross sections from \Eq{EQ_HANG_EE} and those entering the SIDIS cross section from Eqs.~(\ref{EQ_HANG_SIDIS})-(\ref{EQ_HPERP_SIDIS})  are shown together to allow a comparison. We also depict the analyzing powers for $\tilde{G}_1^\perp$ extracted in Ref.~\cite{Matevosyan:2017alv} for comparison. We have explicitly checked that the SIDIS DiFFs computed using the individually extracted moments via Eqs.~(\ref{EQ_HANG_EXTRACT})-(\ref{EQ_HPERP_EXTRACT}) agree with the calculations using Eqs.~(\ref{EQ_HANG_SIDIS_EXTRACT})-(\ref{EQ_HPERP_SIDIS_EXTRACT}). For $N_L=2$, it is important to note, that the magnitudes of $\tilde{H}_1^{\SA ,[0]}$ and $\tilde{H}_1^{\SA, SIDIS}$ are comparable, while the magnitude of $\tilde{H}_1^{\perp, SIDIS}$ is much smaller than that for $\tilde{H}_1^{\perp ,[0]}$, especially for large $z$. This result is easy to understand by recalling those in Fig.~(\ref{PLOT_H_MOMS_NL2}) and  Eqs.~(\ref{EQ_HANG_SIDIS})-(\ref{EQ_HPERP_SIDIS}). $\tilde{H}_1^{\SA,[1]}$ has a much more significant impact on $\tilde{H}_1^{\perp,SIDIS}$ than $\tilde{H}_1^{\perp,[1]}$ on $\tilde{H}_1^{\SA, SIDIS}$ because of the considerable differences in their relative magnitudes. The large differences of the magnitudes and shapes of $\tilde{H}_1^{\SA, SIDIS}$ and $\tilde{H}_1^{\perp,SIDIS}$ for $N_L=6$ have the same origins. For $N_L=6$ results, we observe a notable difference between the analyzing powers of $\tilde{H}_1^{\SA ,[0]}$ and $\tilde{H}_1^{\SA, SIDIS}$, which grow with $z$. Also, it is worthwhile to compare the results for $\tilde{H}_1^{\SA ,[0]}$ and $\tilde{G}_1^\perp$, where the latter is much smaller, especially in the low- to mid-$z$ region. The cuts on the minimum values for the $z$ of each hadron in the pair,  $z_{1,2}\geq 0.1$, do not significantly change the overall conclusions for the results on the analyzing powers. 

\begin{figure}[tb]
\centering 
\subfigure[]
 {
\includegraphics[width=\ImL]{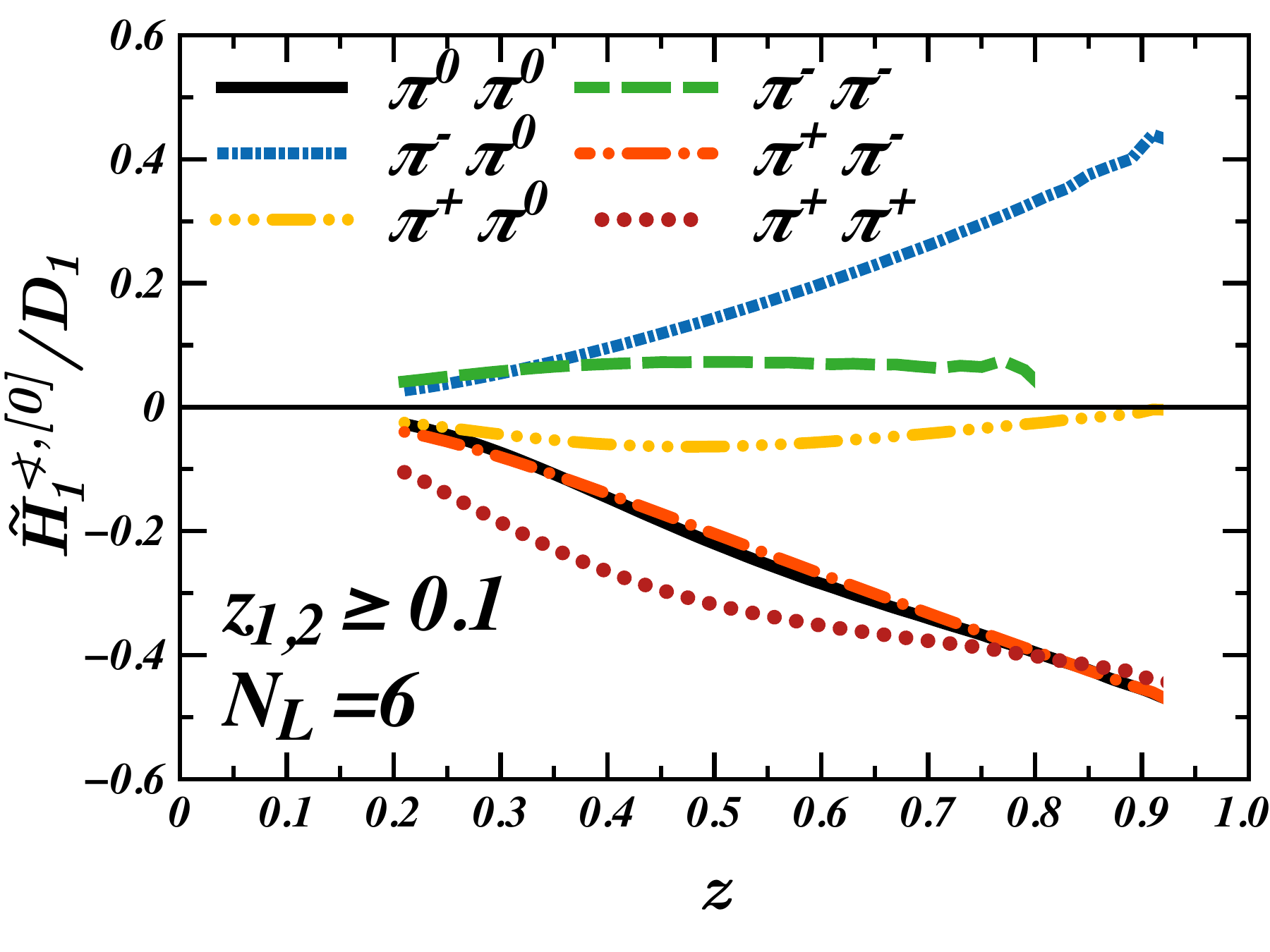}
}
\\ \GapSubf
\subfigure[]
 {
\includegraphics[width=\ImL]{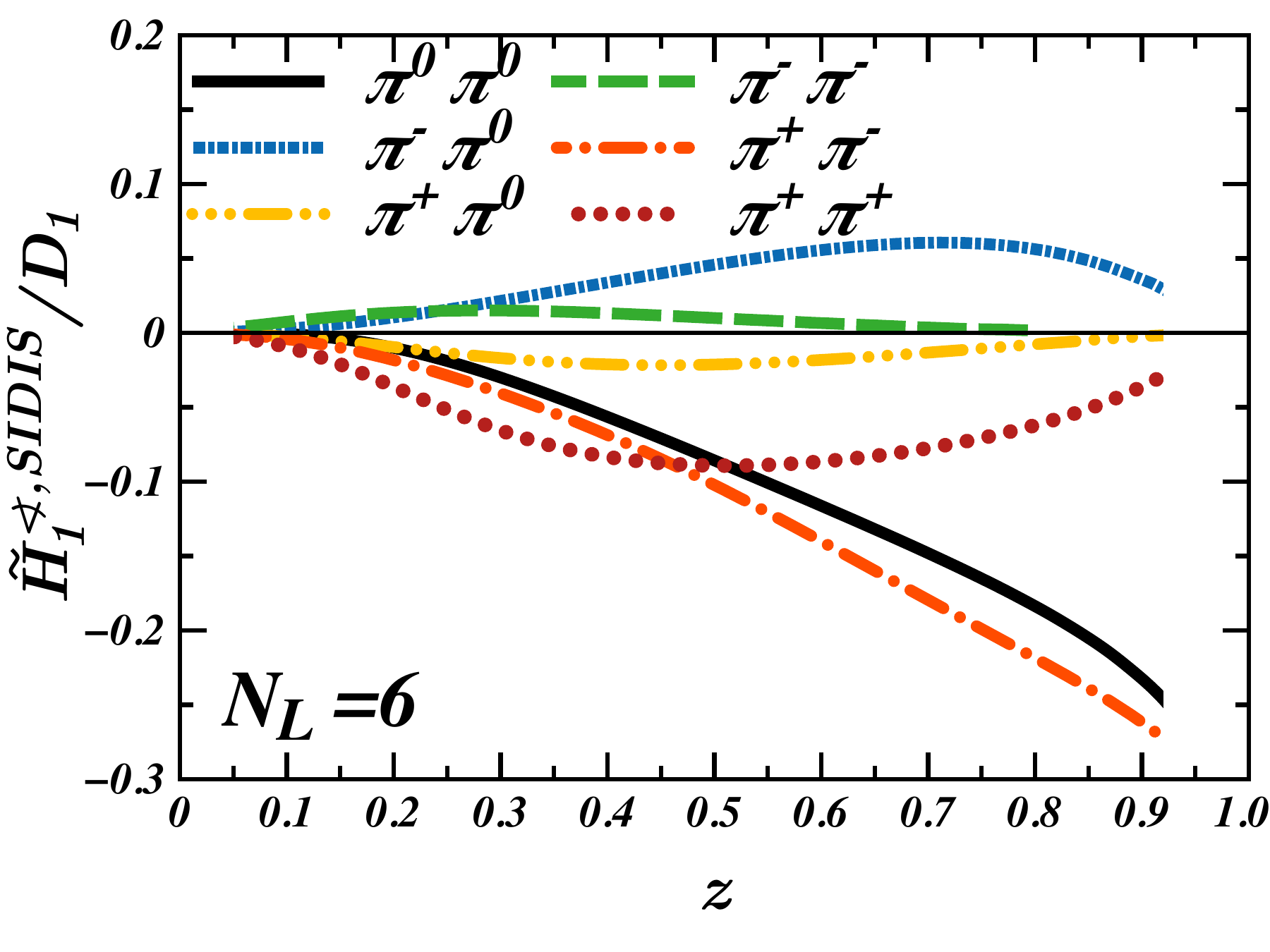}
}
\GapCapt
\caption{The comparison of ratios $\tilde{H}_1^{\SA, [0]}(z)/D_1(z)$ with the cuts (a) $z_{1,2}\geq 0.1$, and (b) $\tilde{H}_1^{\SA, SIDIS}(z)/D_1(z)$, for all the possible pion pairs extracted from MC simulations with $N_L=6$. We used $z$-ordering to assign the first and second hadrons in the same-charged pairs.}
\label{PLOT_HANG_RAT_PAIRS}
\end{figure}

In the MC simulations, we have ready access to all the possible combinations of pion pairs. In Fig.~\ref{PLOT_HANG_RAT_PAIRS}(a), we depict as an example the analyzing power $\tilde{H}_1^{\SA, [0]}(z)/D_1(z)$ for all the pion pairs. Here, similar to our previous work, we again use the $z$-ordering for the same-charged pairs by assigning hadrons the labels first and second, so $z_1 \geq z_2$. We use simulations with  $N_L=6$ and the cut $z_{1,2}\geq 0.1$. The analyzing power for $\pi^+\pi^+$ pairs has the same sign and a larger magnitude  than that for $\pi^+\pi^-$ pairs, while the analyzing power for $\pi^-\pi^0$ pairs has an opposite sign. The results for $\pi^-\pi^-$ pairs are smaller and not well determined for $z\gtrsim 0.8$.  The reason is that a $u$ quark needs to emit at least four hadrons to produce two $\pi^-$ (along with two $\pi^+$), with the first $\pi^-$ appearing at rank $2$ or higher. Thus, the probability of finding  such a pair with a very large value of $z\geq 0.8$ is extremely small.  In Fig.~\ref{PLOT_HANG_RAT_PAIRS}(b), we depict the analogous results for the analyzing power $\tilde{H}_1^{\SA, SIDIS}(z)/D_1(z)$, without the cuts on $z_{1,2}$. Here again the analyzing power for $\pi^+\pi^+$ pairs has the same sign as that for $\pi^+\pi^-$ pairs, but becomes smaller in magnitude for values of $z\gtrsim 0.5$. 

\section{Validation of MC results}
\label{SEC_VALIDATION}

 In this section we derive explicit integral expressions for $H_1^\SA$ and $H_1^\perp$ for hadronization with only two produced hadrons, $N_L=2$. We use these expressions to cross-check the results extracted from the MC simulations, for validation of our method. They are also important in elucidating the underlying mechanism for the generation of these DiFFs via single-hadron SFs.  We follow the same approach as in our previous work in Refs.~\cite{Matevosyan:2016fwi,Matevosyan:2017alv}, where similar expressions were derived for the unpolarized and unfavored Collins FFs and unpolarized and helicity-dependent DiFFs.  We consider the case in which the initial transversely polarized quark $q$  sequentially fragments into two hadrons $h_1$ and $h_2$. We denote the spin vector of the initial quark as  with $\vect{s}_q = (\vect{s}_T, 0 )$, and the spin vector of the  remnant quark  $q_1$ after the emission of the first hadron as $\vect{s}_{q_1} = (\vect{s}_{T_1}, s_{L_1})$. The light-cone momentum fraction and the transverse momentum of $q_1$ with respect to the initial quark $q$ are denoted by $\eta_1$ and $\pe{1}$, while those for $h_2$ with respect to $q_1$ are denoted by $\eta_2$ and $\pe{2}$. Using momentum conservation and a Lorentz boost we can calculate the light-cone momentum fractions $z_1$, $z_2$ and the transverse momenta $\Pe{1}$, $\Pe{2}$ of $h_1$ and $h_2$ with respect to $q$:
\al
{ \label{EQ_H1_Z}
z_1 &= 1-\eta_1,
\\ \label{EQ_H2_Z}
z_2 &= \eta_1 \eta_2,
\\ \label{EQ_H1_TM}
\Pe{1} &= - \pe{1},
\\ \label{EQ_H2_TM}
\Pe{2} &= \pe{2} + \eta_1\pe{1}.
}

We calculate each DiFF by identifying the corresponding term in the polarized number density $F$  using Eq.~(\ref{EQ_F_ANG}).  The number density itself can be easily expressed as a  product of the two number densities for the processes $q\to h_1 + q_1$ and $q_1\to h_2 + q_2$:
\al
{
  \label{EQ_Q_to_2H}
F^{(2)}_{q\to h_1 h_2}(\eta_1, &\pe{1}, \eta_2, \pe{2}; \sq{q} )
\\ \non
=\sum_{q_1}&  \hat{f}^{q\to q_1 }(\eta_1, \pe{1};  \sq{q} )  \cdot \hat{f}^{q_1\to h_2}(\eta_2, \pe{2}; \sq{q_1} ).
}
The elementary probability densities are expressed in terms of the corresponding SFs,
\al
{
\label{EQ_FHAT_Q_Q1}
 \hat{f}^{q\to q_1 }(z, \pe{}; & \vect{s} )
 \\ \non 
 = \hat{D}^{(q\to q_1)}&(z, \psq{})  + \frac{(\pe{}\times \sT) \cdot \hat{\vect{z}} }{z M_{q_1} } \ \hat{H}^{\perp({q\to q_1})}(z,\psq{}),
}
\al
{ 
\label{EQ_FHAT_Q_H}
  \hat{f}^{q \to h}(z, \pe{};& \vect{s} )  
\\ \non
= \hat{D}^{(q\to h)}&(z, \psq{})  + \frac{(\pe{}\times \sT) \cdot \hat{\vect{z}}}{z m_h}\ \hat{H}^{\perp({q\to h})}(z,\psq{}),
}
where $M_{q_1}$ and $m_h$ are the masses of the remnant quark and the produced hadron, respectively.

Thus, we only need to determine the transverse component of the polarization vector of $q_1$. For a transversely polarized initial quark, $q$, this is determined using the corresponding SFs and the momentum~\cite{Matevosyan:2016fwi}
\al
{
\label{EQ_ST1}
 \vect{s}_{T_1}  = \frac{1}{ \hat{f}^{q\to q_1}(\eta_1, \pe{1};  \sq{q}) } 
&\Bigg( \frac{ \pe{1}' }{\eta_1 M_{q_1} }  \hat{D}_T^\perp(\eta_1,\psqn{1})  
 \\ \non
 + \vect{s}_T \hat{H}_T(\eta_1,\psqn{1})&
 + \frac{\pe{1} (\pe{1} \cdot \vect{s}_T)}{\eta_1^2 M_{q_1}^2} \hat{H}_T^\perp(\eta_1,\psqn{1}) 
  \Bigg),
}
where $\hat{D}_T^\perp$, $\hat{H}_T$, and $\hat{H}_T^\perp$ are the polarizing, transversity, and pretzelocity transverse momentum dependent SFs. The transverse vector $\pe{1}'$ is defined as $\pe{1}' \equiv (-p_{1y},p_{1x})$. 

 We can then easily derive the integral expressions for the $z$ dependence of the two DiFFs in which we are interested. Here we will only describe the relevant terms contributing ${H}_1^{\SA}$ and $ {H}_1^{\perp}$ to avoid explicitly writing out lengthy expressions
\al
{
\label{EQ_HANG_NL2}
&\frac{H_1^{\SA (2)}(z, \xi, \kT^2, \RT^2, \kT \cdot \RT) }{M_1 + M_2} 
 =\frac{z^3}{1-z\xi}
 \\ \non
& \times \Bigg[
 \hat{H}^{\perp (q\to q_1)} \hat{D}^{(q_1\to h_2)}
 + \frac{1-z}{1-z\xi} \hat{H}_T^{(q\to q_1)} \hat{H}^{\perp ({q_1\to h_2})}
\\ \non
& \quad \quad -z \Big( (\kT \cdot \RT) - z \xi \ \kT^2 \Big) 
\hat{H}_T^{\perp (q\to q_1)} \hat{H}^{\perp({q_1\to h_2})}
 \Bigg],
}
\al
{
\label{EQ_HPERP_NL2}
&\frac{ H_1^{\perp(2)}(z, \xi, \kT^2, \RT^2, \kT \cdot \RT) }{M_1 + M_2} 
= \frac{z^4}{1-z\xi}
 \\ \non
& \times \Bigg[
- \xi  \hat{H}^{\perp (q\to q_1)} \hat{D}^{(q_1\to h_2)}
+ \frac{1-\xi}{1-z\xi} \hat{H}_T^{(q\to q_1)} \hat{H}^{\perp ({q_1\to h_2})}
\\ \non
& \quad \quad - \Big(z \xi \ (\kT \cdot \RT) - \RT^2  \Big)  
\hat{H}_T^{\perp (q\to q_1)} \hat{H}^{\perp({q_1\to h_2})}
\Bigg],
}
where the SFs for $q\to q_1$ are functions of $(\eta_1, \pe{1})$; those for $q_1 \to h_2$ depend on $(\eta_2, \pe{2})$; and we have redefined the SFs to absorb the corresponding denominators $(\eta_{1}\ M_{q_1})$,$(\eta_{1}\ M_{q_1})^{2}$, $(\eta_{2}\ m_{h_2})$ appearing in  Eqs.~(\ref{EQ_FHAT_Q_Q1})-(\ref{EQ_ST1}). 

 We can readily see from Eqs.~(\ref{EQ_HANG_NL2},\ref{EQ_HPERP_NL2}) that both of the transverse-polarization-dependent DiFFs share the same generation mechanism. The first contribution, $\hat{H}^{\perp (q\to q_1)} \hat{D}^{(q_1\to h_2)}$, encodes the transverse momentum "recoil" effect in the hadronization chain. The Collins effect in the first emission step creates some modulation with respect to the azimuthal angle of the transverse momentum of the remnant quark, and consequently $h_1$ via momentum conservation (\ref{EQ_H1_TM}). A fraction of this momentum is then transferred to the second hadron $h_2$ according to Eq.~(\ref{EQ_H2_TM}).  The second and third terms, $ \hat{H}_T^{(q\to q_1)} \hat{H}^{\perp ({q_1\to h_2})}$ and $ \hat{H}_T^{\perp(q\to q_1)} \hat{H}^{\perp ({q_1\to h_2})}$, involve the transversity and the pretzelocity SFs in the first emission step, which transfer the transverse polarization of the initial quark to $q_1$; see \Eq{EQ_ST1}.  Then, this transverse polarization is correlated with the transverse momentum of the produced hadron, $h_2$,  via the  Collins  effect in the second emission step. 

 Explicit integral expressions (IE) for $H_1^{\SA ,[0]}$, $H_1^{\SA, SIDIS}$, $H_1^{\perp ,[0]}$, and $H_1^{\perp, SIDIS}$ are easily obtained by calculating the Fourier cosine moments using  Eqs.~(\ref{EQ_HANG_MOM_MH})-(\ref{EQ_HPERP_MOM_MH}) and forming the corresponding combinations given in  Eqs.~(\ref{EQ_HANG_EE})-(\ref{EQ_HPERP_SIDIS}). We can then use the NJL model expressions for the SFs from Ref.~\cite{Matevosyan:2016fwi} to compute the integrals and obtain the numerical results. 

\begin{figure}[tb]
\centering 
\subfigure[]
 {
\includegraphics[width=\ImL]{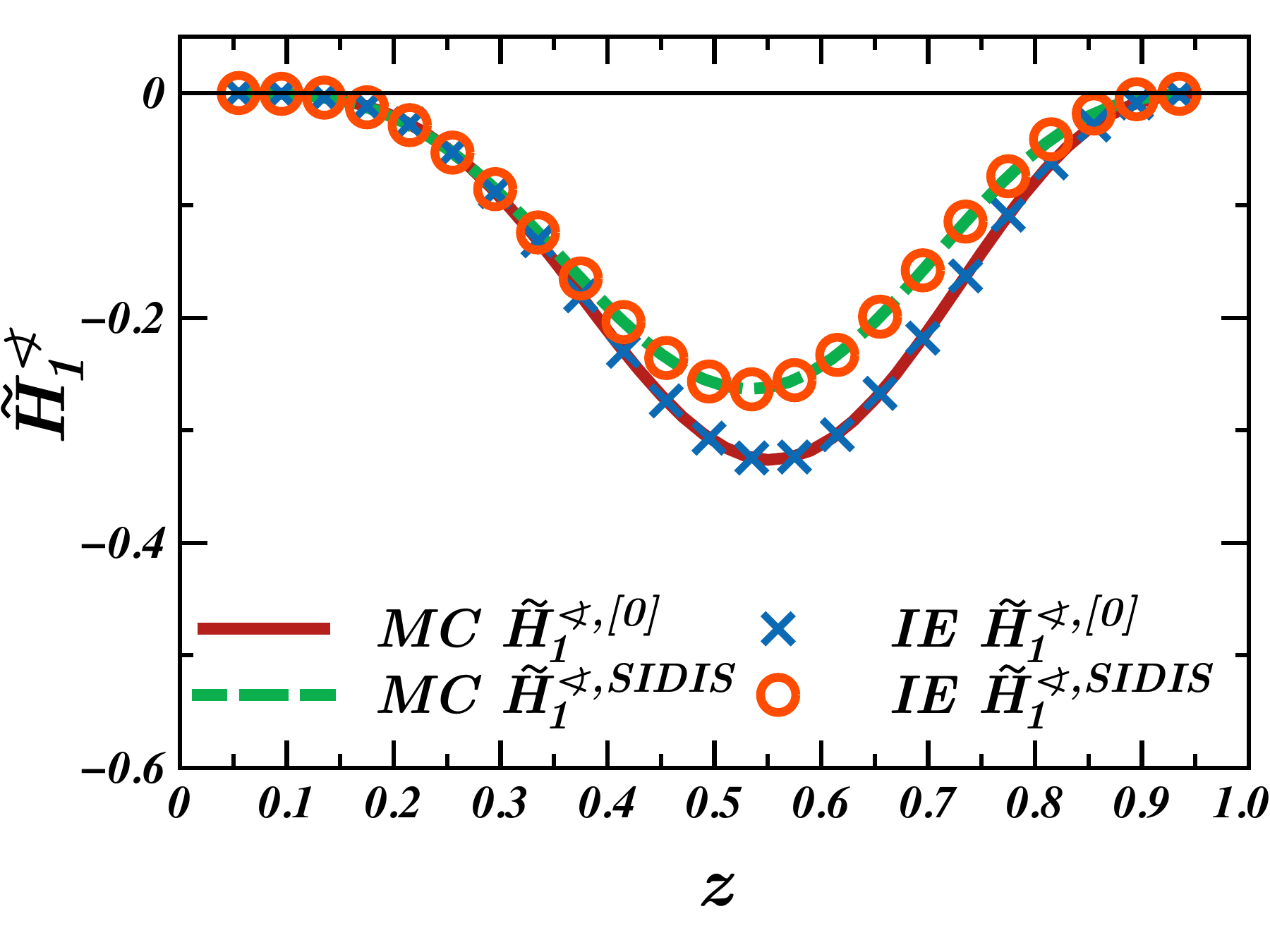}
}
\\ \GapSubf
\subfigure[]
 {
\includegraphics[width=\ImL]{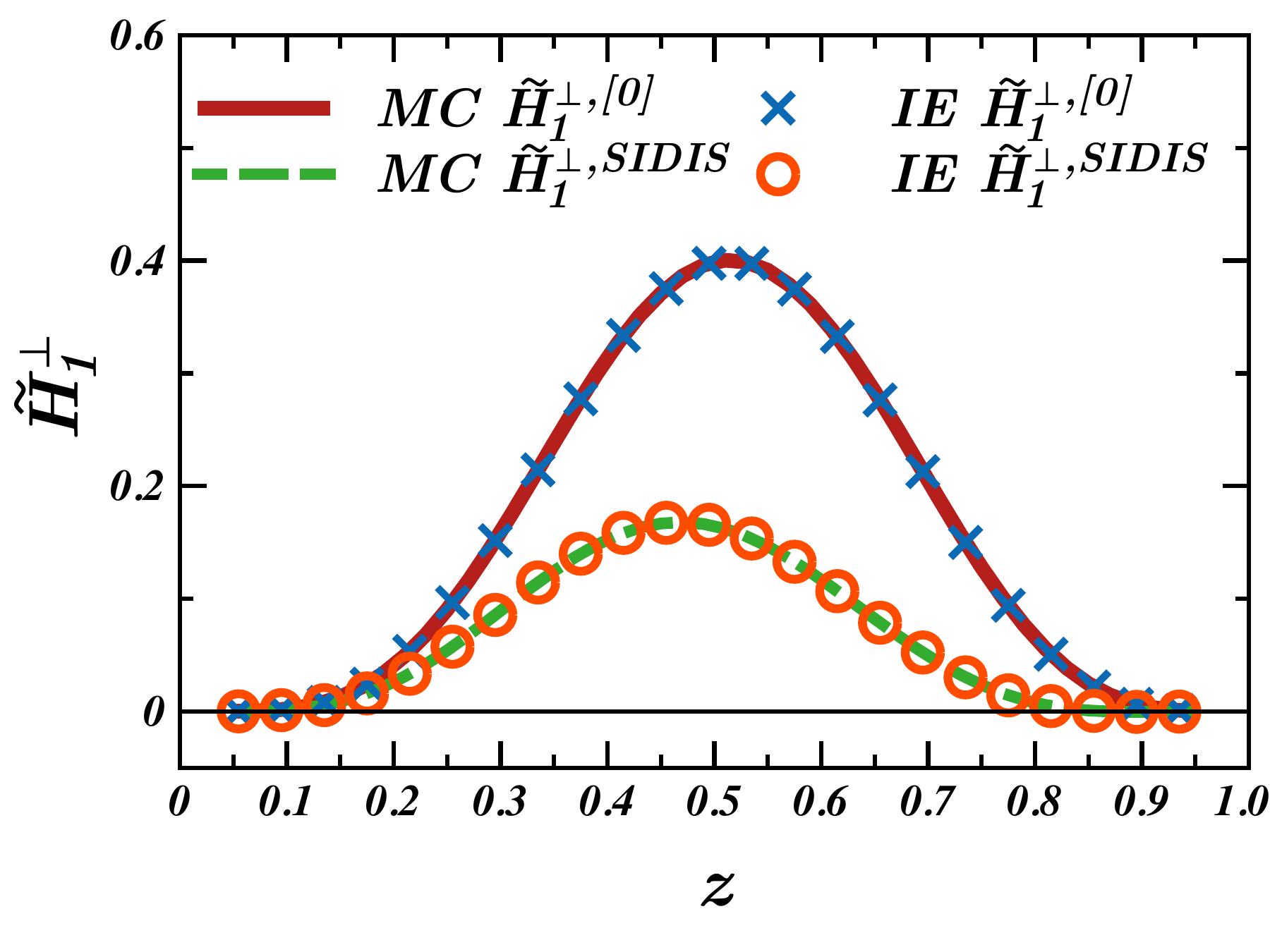}
}
\GapCapt
\caption{Comparison of the $\pi^+\pi^-$ results for (a)  $\tilde{H}_1^{\SA ,[0]}(z)$, $\tilde{H}_1^{\SA, SIDIS}(z)$ and (b) $\tilde{H}_1^{\perp ,[0]}$, $\tilde{H}_1^{\perp, SIDIS}$, obtained using IEs and MC simulations for $N_L=2$.}
\label{PLOT_HANG_HPERP_NL2}
\end{figure}

In Fig.~\ref{PLOT_HANG_HPERP_NL2} we compare the transverse-polarization-dependent DiFFs calculated using the IEs and MC simulations for $\pi^+\pi^-$ pairs, in particular the results for (a) $\tilde{H}_1^{\SA ,[0]}$, $\tilde{H}_1^{\SA, SIDIS}$ and (b) $\tilde{H}_1^{\perp ,[0]}$, $\tilde{H}_1^{\perp, SIDIS}$. The MC simulations are plotted using the solid and dashed lines, while the crosses and open circles show the IE results. We achieved excellent agreement between the two methods for all four DiFFs considered, demonstrating the reliability of our MC simulations.

\section{Conclusions}
\label{SEC_CONCLUSIONS}

 In this paper we studied the two DiFFs describing the correlations between the transverse polarization of the fragmenting quark and the transverse momenta of a produced hadron pair in the hadronization process. We first briefly overviewed the two-hadron fragmentation kinematics and the field-theoretical definitions of the leading-twist DiFFs in Sec.~\ref{SEC_DIFF_FORM}.  The Fourier decomposition of these DiFFs only contains the cosine moments that were defined in Eqs.~(\ref{EQ_D1_MOM_MH})-(\ref{EQ_HPERP_MOM_MH}). 

An important observation here is that there is a discrepancy between the definitions of the integrated DiFFs that enter into the cross section expression for $e^+e^-$ production of two back-to-back hadron pairs in Ref.~\cite{Boer:2003ya} and the production of hadron pairs in SIDIS in Ref.~\cite{Bacchetta:2003vn}. The $e^+e^-$ expression shown in \Eq{EQ_HANG_EE} contains the zeroth cosine moment of the unintegrated DiFF, while those for SIDIS in Eqs.~(\ref{EQ_HANG_SIDIS})-(\ref{EQ_HPERP_SIDIS}) also contain the first cosine moments of the alternate DiFFs. This difference can potentially have a significant impact on the combined experimental and phenomenological efforts~\cite{Bacchetta:2011ip, Bacchetta:2012ty, Courtoy:2012ry,Pisano:2015wnq, Radici:2015mwa} of studying the nucleon transversity parton distribution functions by analyzing the two-hadron SIDIS measurements~\cite{Airapetian:2008sk,Adolph:2012nw,Adolph:2014fjw}. The crucial ingredient here is  the $H_1^\SA$ IFF, that is obtained by fitting the measured $e^+e^-$  asymmetries~\cite{Vossen:2011fk}. Of course, a thorough review of the cross section derivations for theis process is needed to ensure against the possibility that terms were missed in the initial derivations in Ref.~\cite{Boer:2003ya}  that might possibly resolve the discrepancies. We could not find any omissions in deriving the asymmetries from the corresponding cross sections, but a full, systematic review of the derivations is beyond the scope of this work.  On the other hand, our aim has been to use the quark-jet model to give a quantitative estimate of the differences between $H_1^{\SA ,[0]}$ and  $H_1^{\SA,SIDIS}$. 

In the MC simulations, we directly calculated the various number densities, written out in \Eq{EQ_F_VEC} in Sec.~\ref{SEC_MC}. Then, we derived the method of extracting the cosine moments of the DiFFs from the polarized number densities. A straightforward prescription~(\ref{EQ_HANG_SIDIS_EXTRACT})-(\ref{EQ_HPERP_SIDIS_EXTRACT}) for computing the SIDIS DiFFs directly from the number density was described. This provides an alternative to adding the cosine moments in Eqs.~(\ref{EQ_HANG_SIDIS})-(\ref{EQ_HPERP_SIDIS}), obtained using Eqs.~(\ref{EQ_HANG_EXTRACT})-(\ref{EQ_HPERP_EXTRACT}). This prescription also outlines how these quantities appear in the cross section of the integrated SIDIS cross section, first demonstrated in Ref.~\cite{Bacchetta:2003vn}.  For example, when the number density (\ref{EQ_F_ANG}) (or the quark-quark-correlator in~\cite{Bacchetta:2003vn}) is integrated over $\fK$, then the $H_1^\perp$ term does not completely disappear. Rather, its first cosine moment survives, multiplied by $\sin(\fR-\fs)$.
  
  The numerical results for the $z$ dependence of $\tilde{H}_1^\SA$ and $\tilde{H}_1^\perp$ for pion pairs were presented in Sec.~\ref{SEC_NJL_JET}. The plots for their first two moments for $\pi^+\pi^-$ pairs in MC simulations with $N_L=2$ were presented in Fig.~\ref{PLOT_H_MOMS_NL2}, demonstrating the opposite signs for $\tilde{H}_1^\SA$ and $\tilde{H}_1^\perp$. Moreover, the first moment of $\tilde{H}_1^{\perp,[1]}$ is significantly smaller in magnitude than  $\tilde{H}_1^{\SA,[0]}$, while  $\tilde{H}_1^{\SA,[1]}$ is comparable in magnitude to  $\tilde{H}_1^{\perp,[0]}$. This has important implications for the combinations $\tilde{H}_1^{\SA ,[0]}$,  $\tilde{H}_1^{\SA,SIDIS}$, $\tilde{H}_1^{\perp ,[0]}$, and  $\tilde{H}_1^{\perp,SIDIS}$. Their analyzing powers are depicted in Fig.~\ref{PLOT_H_RAT_PAIRS} for $N_L=2$, $N_L=6$, and $N_L=6$ with a $z$-cut, alongside the results for $\tilde{G}_1^\perp$ computed in Ref.~\cite{Matevosyan:2017alv}. The rapid convergence of the analyzing power of $\tilde{H}_1^{\perp,[0]}(z)$ with increasing $N_L$ was shown in Fig.~\ref{PLOT_D1_NLX}. Examining the $N_L=6$ results, it is apparent that  $\tilde{H}_1^{\SA ,[0]}$ and  $\tilde{H}_1^{\perp ,[0]}$ are similar in magnitude and opposite in sign. The suppression of $\tilde{H}_1^{\perp ,SIDIS}$ can be understood from the earlier observations about the relative sizes of the first two moments of the DiFFs. 

We observe that the magnitude of  $\tilde{H}_1^{\SA ,SIDIS}$ is roughly $50\%$ smaller than the magnitude of $\tilde{H}_1^{\SA ,[0]}$ for $z>0.2$. Further, the magnitude of $\tilde{G}_1^{\perp}$ is significantly suppressed compared to $\tilde{H}_1^{\SA ,[0]}$ and $\tilde{H}_1^{\SA ,SIDIS}$ for the majority of the $z$ region. This further strengthens the conclusions drawn in Ref.~\cite{Matevosyan:2017alv}, supporting the nonobservation of the $\tilde{G}_1^\perp$ signal at BELLE~\cite{Abdesselam:2015nxn, Vossen:2015znm} and COMPASS~\cite{Sirtl:2017rhi}, while the corresponding asymmetries for IFF have been successfully measured in both experiments~\cite{Vossen:2011fk,Adolph:2012nw,Adolph:2014fjw}. The  $\tilde{H}_1^{\perp ,SIDIS}$ appears to be significantly smaller in magnitude than $\tilde{H}_1^{\SA ,SIDIS}$; thus the corresponding signal in the SIDIS cross section should be also suppressed. The minimum $z$-cut on each hadron in the pair produced no qualitatively significant change in the results. The last result presented in the section was for the analyzing power of $\tilde{H}_1^{\SA ,[0]}$ and $\tilde{H}_1^{\SA ,SIDIS}$ for all the possible pion pairs in Fig.~\ref{PLOT_HANG_RAT_PAIRS}, showing that the $\tilde{H}_1^{\SA ,[0]}$ signal for the $\pi^+\pi^+$ pairs has the same sign and larger magnitude than the signal for the $\pi^+\pi^-$ pairs. On the other hand, the $\tilde{H}_1^{\SA ,SIDIS}$ signal for the $\pi^+\pi^+$ becomes smaller in magnitude than that for $\pi^+\pi^-$ pairs for the large values of $z\gtrsim 0.5$. 
   
 Finally, we performed a validation test of our MC extraction of the DiFFs in Sec.~\ref{SEC_VALIDATION} by explicitly calculating the polarized number density for two-hadron emission and identifying the terms corresponding to the different DiFFs. Then, the numerical values were computed by substituting the NJL-model-calculated SFs and integrating over the transverse momenta. The results for both $e^+e^-$ and SIDIS combinations are illustrated in Fig.~\ref{PLOT_HANG_HPERP_NL2}, showing perfect agreement with the MC simulations. In addition, in the results for the unintegrated  ${H}_1^\SA$ and ${H}_1^\perp$ in Eqs~(\ref{EQ_HANG_NL2})-(\ref{EQ_HPERP_NL2}), we can readily identify the mechanism for their generation via the single-hadron-emission elementary SFs. There are the three contributions involving the same SFs for both DiFFs. The "recoil" transverse momentum modulated by the Collins effect in the first hadron emission and the two terms describing the transfer of the transverse polarization of the quark to the remnant quark after the first emission. In all of the terms, the Collins function is present, generating the correlations between transverse polarization and the transverse momenta. These results are very similar to those for the Collins function for rank-2 hadrons derived in Sec.~II.C of Ref.~\cite{Matevosyan:2016fwi}.
 
 It is important to note several caveats. In this work we used the low scale effective NJL model input for the numerical computations. Thus, QCD evolution of these DiFFs~\cite{Ceccopieri:2007ip} needs to be performed before comparing with the experiment. We roughly modeled the impact of QCD evolution by employing the $(1-z)^4$ ansatz for the input SFs to mimic the effects of evolution by shifting them to a lower $z$ region. Nevertheless, a proper evolution of the results will be performed in our future work, as done previously for the unpolarized DiFFs in Ref.~\cite{Matevosyan:2013aka}. Another significant omission is the production of kaons and the vector mesons, as well as the strong decays of the vector resonances. The rich structure of the invariant mass dependence of the unpolarized DiFF is generated by the decays of these resonances, as shown both in MC simulations~\cite{Bacchetta:2006un,Matevosyan:2013aka} and the recent BELLE results~\cite{Seidl:2017qhp}. Consequently, any possible two-hadron interference effects between the decay products of the resonances, conventionally attributed to the generation of the IFF (see, e.g.~\cite{Bacchetta:2006un}), were also omitted. 
 
 In summary,  in the series of works including Ref.~\cite{Matevosyan:2017alv} and this manuscript, we successfully calculated all four leading-twist DiFFs for pion pairs within the quark-jet hadronization framework. We made several important observations, most notably about the relative sizes of the various DiFFs and their generation mechanisms. Future work will be aimed at refining the model, by including, for example, several hadron production, strong decay channels and QCD evolution.
 
\vspace{10pt}
\section*{Acknowledgements}

 The work of H.H.M. and A.W.T. was supported by the Australian Research Council through the ARC Centre of Excellence for Particle Physics at the Terascale (CE110001104), and by an ARC Australian Laureate Fellowship FL0992247 and Discovery Project No. DP151103101, as well as by the University of Adelaide. A.K. was supported by the A.I. Alikhanyan National Science Laboratory (YerPhI) Foundation, Yerevan, Armenia.


\bibliographystyle{apsrev4-1}
\bibliography{fragment}

\begin{thebibliography}{36}%
\makeatletter
\providecommand \@ifxundefined [1]{%
 \@ifx{#1\undefined}
}%
\providecommand \@ifnum [1]{%
 \ifnum #1\expandafter \@firstoftwo
 \else \expandafter \@secondoftwo
 \fi
}%
\providecommand \@ifx [1]{%
 \ifx #1\expandafter \@firstoftwo
 \else \expandafter \@secondoftwo
 \fi
}%
\providecommand \natexlab [1]{#1}%
\providecommand \enquote  [1]{``#1''}%
\providecommand \bibnamefont  [1]{#1}%
\providecommand \bibfnamefont [1]{#1}%
\providecommand \citenamefont [1]{#1}%
\providecommand \href@noop [0]{\@secondoftwo}%
\providecommand \href [0]{\begingroup \@sanitize@url \@href}%
\providecommand \@href[1]{\@@startlink{#1}\@@href}%
\providecommand \@@href[1]{\endgroup#1\@@endlink}%
\providecommand \@sanitize@url [0]{\catcode `\\12\catcode `\$12\catcode
  `\&12\catcode `\#12\catcode `\^12\catcode `\_12\catcode `\%12\relax}%
\providecommand \@@startlink[1]{}%
\providecommand \@@endlink[0]{}%
\providecommand \url  [0]{\begingroup\@sanitize@url \@url }%
\providecommand \@url [1]{\endgroup\@href {#1}{\urlprefix }}%
\providecommand \urlprefix  [0]{URL }%
\providecommand \Eprint [0]{\href }%
\providecommand \doibase [0]{http://dx.doi.org/}%
\providecommand \selectlanguage [0]{\@gobble}%
\providecommand \bibinfo  [0]{\@secondoftwo}%
\providecommand \bibfield  [0]{\@secondoftwo}%
\providecommand \translation [1]{[#1]}%
\providecommand \BibitemOpen [0]{}%
\providecommand \bibitemStop [0]{}%
\providecommand \bibitemNoStop [0]{.\EOS\space}%
\providecommand \EOS [0]{\spacefactor3000\relax}%
\providecommand \BibitemShut  [1]{\csname bibitem#1\endcsname}%
\let\auto@bib@innerbib\@empty
\bibitem [{\citenamefont {Bianconi}\ \emph
  {et~al.}(2000{\natexlab{a}})\citenamefont {Bianconi}, \citenamefont {Boffi},
  \citenamefont {Jakob},\ and\ \citenamefont {Radici}}]{Bianconi:1999cd}%
  \BibitemOpen
  \bibfield  {author} {\bibinfo {author} {\bibfnamefont {A.}~\bibnamefont
  {Bianconi}}, \bibinfo {author} {\bibfnamefont {S.}~\bibnamefont {Boffi}},
  \bibinfo {author} {\bibfnamefont {R.}~\bibnamefont {Jakob}}, \ and\ \bibinfo
  {author} {\bibfnamefont {M.}~\bibnamefont {Radici}},\ }\href {\doibase
  10.1103/PhysRevD.62.034008} {\bibfield  {journal} {\bibinfo  {journal}
  {Phys.Rev.}\ }\textbf {\bibinfo {volume} {D62}},\ \bibinfo {pages} {034008}
  (\bibinfo {year} {2000}{\natexlab{a}})},\ \Eprint
  {http://arxiv.org/abs/hep-ph/9907475} {arXiv:hep-ph/9907475 [hep-ph]}
  \BibitemShut {NoStop}%
\bibitem [{\citenamefont {Bianconi}\ \emph
  {et~al.}(2000{\natexlab{b}})\citenamefont {Bianconi}, \citenamefont {Boffi},
  \citenamefont {Jakob},\ and\ \citenamefont {Radici}}]{Bianconi:1999uc}%
  \BibitemOpen
  \bibfield  {author} {\bibinfo {author} {\bibfnamefont {A.}~\bibnamefont
  {Bianconi}}, \bibinfo {author} {\bibfnamefont {S.}~\bibnamefont {Boffi}},
  \bibinfo {author} {\bibfnamefont {R.}~\bibnamefont {Jakob}}, \ and\ \bibinfo
  {author} {\bibfnamefont {M.}~\bibnamefont {Radici}},\ }\href {\doibase
  10.1103/PhysRevD.62.034009} {\bibfield  {journal} {\bibinfo  {journal}
  {Phys.Rev.}\ }\textbf {\bibinfo {volume} {D62}},\ \bibinfo {pages} {034009}
  (\bibinfo {year} {2000}{\natexlab{b}})},\ \Eprint
  {http://arxiv.org/abs/hep-ph/9907488} {arXiv:hep-ph/9907488 [hep-ph]}
  \BibitemShut {NoStop}%
\bibitem [{\citenamefont {Radici}\ \emph {et~al.}(2002)\citenamefont {Radici},
  \citenamefont {Jakob},\ and\ \citenamefont {Bianconi}}]{Radici:2001na}%
  \BibitemOpen
  \bibfield  {author} {\bibinfo {author} {\bibfnamefont {M.}~\bibnamefont
  {Radici}}, \bibinfo {author} {\bibfnamefont {R.}~\bibnamefont {Jakob}}, \
  and\ \bibinfo {author} {\bibfnamefont {A.}~\bibnamefont {Bianconi}},\ }\href
  {\doibase 10.1103/PhysRevD.65.074031} {\bibfield  {journal} {\bibinfo
  {journal} {Phys.Rev.}\ }\textbf {\bibinfo {volume} {D65}},\ \bibinfo {pages}
  {074031} (\bibinfo {year} {2002})},\ \Eprint
  {http://arxiv.org/abs/hep-ph/0110252} {arXiv:hep-ph/0110252 [hep-ph]}
  \BibitemShut {NoStop}%
\bibitem [{\citenamefont {Bacchetta}\ and\ \citenamefont
  {Radici}(2004)}]{Bacchetta:2003vn}%
  \BibitemOpen
  \bibfield  {author} {\bibinfo {author} {\bibfnamefont {A.}~\bibnamefont
  {Bacchetta}}\ and\ \bibinfo {author} {\bibfnamefont {M.}~\bibnamefont
  {Radici}},\ }\href {\doibase 10.1103/PhysRevD.69.074026} {\bibfield
  {journal} {\bibinfo  {journal} {Phys.Rev.}\ }\textbf {\bibinfo {volume}
  {D69}},\ \bibinfo {pages} {074026} (\bibinfo {year} {2004})},\ \Eprint
  {http://arxiv.org/abs/hep-ph/0311173} {arXiv:hep-ph/0311173 [hep-ph]}
  \BibitemShut {NoStop}%
\bibitem [{\citenamefont {Bacchetta}\ \emph {et~al.}(2011)\citenamefont
  {Bacchetta}, \citenamefont {Courtoy},\ and\ \citenamefont
  {Radici}}]{Bacchetta:2011ip}%
  \BibitemOpen
  \bibfield  {author} {\bibinfo {author} {\bibfnamefont {A.}~\bibnamefont
  {Bacchetta}}, \bibinfo {author} {\bibfnamefont {A.}~\bibnamefont {Courtoy}},
  \ and\ \bibinfo {author} {\bibfnamefont {M.}~\bibnamefont {Radici}},\ }\href
  {\doibase 10.1103/PhysRevLett.107.012001} {\bibfield  {journal} {\bibinfo
  {journal} {Phys.Rev.Lett.}\ }\textbf {\bibinfo {volume} {107}},\ \bibinfo
  {pages} {012001} (\bibinfo {year} {2011})},\ \Eprint
  {http://arxiv.org/abs/1104.3855} {arXiv:1104.3855 [hep-ph]} \BibitemShut
  {NoStop}%
\bibitem [{\citenamefont {Bacchetta}\ \emph {et~al.}(2013)\citenamefont
  {Bacchetta}, \citenamefont {Courtoy},\ and\ \citenamefont
  {Radici}}]{Bacchetta:2012ty}%
  \BibitemOpen
  \bibfield  {author} {\bibinfo {author} {\bibfnamefont {A.}~\bibnamefont
  {Bacchetta}}, \bibinfo {author} {\bibfnamefont {A.}~\bibnamefont {Courtoy}},
  \ and\ \bibinfo {author} {\bibfnamefont {M.}~\bibnamefont {Radici}},\ }\href
  {\doibase 10.1007/JHEP03(2013)119} {\bibfield  {journal} {\bibinfo  {journal}
  {JHEP}\ }\textbf {\bibinfo {volume} {1303}},\ \bibinfo {pages} {119}
  (\bibinfo {year} {2013})},\ \Eprint {http://arxiv.org/abs/1212.3568}
  {arXiv:1212.3568 [hep-ph]} \BibitemShut {NoStop}%
\bibitem [{\citenamefont {Pisano}\ and\ \citenamefont
  {Radici}(2016)}]{Pisano:2015wnq}%
  \BibitemOpen
  \bibfield  {author} {\bibinfo {author} {\bibfnamefont {S.}~\bibnamefont
  {Pisano}}\ and\ \bibinfo {author} {\bibfnamefont {M.}~\bibnamefont
  {Radici}},\ }\href {\doibase 10.1140/epja/i2016-16155-5} {\bibfield
  {journal} {\bibinfo  {journal} {Eur. Phys. J.}\ }\textbf {\bibinfo {volume}
  {A52}},\ \bibinfo {pages} {155} (\bibinfo {year} {2016})},\ \Eprint
  {http://arxiv.org/abs/1511.03220} {arXiv:1511.03220 [hep-ph]} \BibitemShut
  {NoStop}%
\bibitem [{\citenamefont {Courtoy}\ \emph {et~al.}(2012)\citenamefont
  {Courtoy}, \citenamefont {Bacchetta}, \citenamefont {Radici},\ and\
  \citenamefont {Bianconi}}]{Courtoy:2012ry}%
  \BibitemOpen
  \bibfield  {author} {\bibinfo {author} {\bibfnamefont {A.}~\bibnamefont
  {Courtoy}}, \bibinfo {author} {\bibfnamefont {A.}~\bibnamefont {Bacchetta}},
  \bibinfo {author} {\bibfnamefont {M.}~\bibnamefont {Radici}}, \ and\ \bibinfo
  {author} {\bibfnamefont {A.}~\bibnamefont {Bianconi}},\ }\href {\doibase
  10.1103/PhysRevD.85.114023} {\bibfield  {journal} {\bibinfo  {journal}
  {Phys.Rev.}\ }\textbf {\bibinfo {volume} {D85}},\ \bibinfo {pages} {114023}
  (\bibinfo {year} {2012})},\ \Eprint {http://arxiv.org/abs/1202.0323}
  {arXiv:1202.0323 [hep-ph]} \BibitemShut {NoStop}%
\bibitem [{\citenamefont {Radici}\ \emph {et~al.}(2015)\citenamefont {Radici},
  \citenamefont {Courtoy}, \citenamefont {Bacchetta},\ and\ \citenamefont
  {Guagnelli}}]{Radici:2015mwa}%
  \BibitemOpen
  \bibfield  {author} {\bibinfo {author} {\bibfnamefont {M.}~\bibnamefont
  {Radici}}, \bibinfo {author} {\bibfnamefont {A.}~\bibnamefont {Courtoy}},
  \bibinfo {author} {\bibfnamefont {A.}~\bibnamefont {Bacchetta}}, \ and\
  \bibinfo {author} {\bibfnamefont {M.}~\bibnamefont {Guagnelli}},\ }\href
  {\doibase 10.1007/JHEP05(2015)123} {\bibfield  {journal} {\bibinfo  {journal}
  {JHEP}\ }\textbf {\bibinfo {volume} {05}},\ \bibinfo {pages} {123} (\bibinfo
  {year} {2015})},\ \Eprint {http://arxiv.org/abs/1503.03495} {arXiv:1503.03495
  [hep-ph]} \BibitemShut {NoStop}%
\bibitem [{\citenamefont {Vossen}\ \emph {et~al.}(2011)\citenamefont {Vossen}
  \emph {et~al.}}]{Vossen:2011fk}%
  \BibitemOpen
  \bibfield  {author} {\bibinfo {author} {\bibfnamefont {A.}~\bibnamefont
  {Vossen}} \emph {et~al.} (\bibinfo {collaboration} {Belle Collaboration}),\
  }\href {\doibase 10.1103/PhysRevLett.107.072004} {\bibfield  {journal}
  {\bibinfo  {journal} {Phys.Rev.Lett.}\ }\textbf {\bibinfo {volume} {107}},\
  \bibinfo {pages} {072004} (\bibinfo {year} {2011})},\ \Eprint
  {http://arxiv.org/abs/1104.2425} {arXiv:1104.2425 [hep-ex]} \BibitemShut
  {NoStop}%
\bibitem [{\citenamefont {Airapetian}\ \emph {et~al.}(2008)\citenamefont
  {Airapetian} \emph {et~al.}}]{Airapetian:2008sk}%
  \BibitemOpen
  \bibfield  {author} {\bibinfo {author} {\bibfnamefont {A.}~\bibnamefont
  {Airapetian}} \emph {et~al.} (\bibinfo {collaboration} {HERMES
  Collaboration}),\ }\href {\doibase 10.1088/1126-6708/2008/06/017} {\bibfield
  {journal} {\bibinfo  {journal} {JHEP}\ }\textbf {\bibinfo {volume} {06}},\
  \bibinfo {pages} {017} (\bibinfo {year} {2008})},\ \Eprint
  {http://arxiv.org/abs/0803.2367} {arXiv:0803.2367 [hep-ex]} \BibitemShut
  {NoStop}%
\bibitem [{\citenamefont {Adolph}\ \emph {et~al.}(2012)\citenamefont {Adolph}
  \emph {et~al.}}]{Adolph:2012nw}%
  \BibitemOpen
  \bibfield  {author} {\bibinfo {author} {\bibfnamefont {C.}~\bibnamefont
  {Adolph}} \emph {et~al.} (\bibinfo {collaboration} {COMPASS Collaboration}),\
  }\href {\doibase 10.1016/j.physletb.2012.05.015} {\bibfield  {journal}
  {\bibinfo  {journal} {Phys.Lett.}\ }\textbf {\bibinfo {volume} {B713}},\
  \bibinfo {pages} {10} (\bibinfo {year} {2012})},\ \Eprint
  {http://arxiv.org/abs/1202.6150} {arXiv:1202.6150 [hep-ex]} \BibitemShut
  {NoStop}%
\bibitem [{\citenamefont {Adolph}\ \emph {et~al.}(2014)\citenamefont {Adolph}
  \emph {et~al.}}]{Adolph:2014fjw}%
  \BibitemOpen
  \bibfield  {author} {\bibinfo {author} {\bibfnamefont {C.}~\bibnamefont
  {Adolph}} \emph {et~al.} (\bibinfo {collaboration} {COMPASS Collaboration}),\
  }\href {\doibase 10.1016/j.physletb.2014.06.080} {\bibfield  {journal}
  {\bibinfo  {journal} {Phys.Lett.}\ }\textbf {\bibinfo {volume} {B736}},\
  \bibinfo {pages} {124} (\bibinfo {year} {2014})},\ \Eprint
  {http://arxiv.org/abs/1401.7873} {arXiv:1401.7873 [hep-ex]} \BibitemShut
  {NoStop}%
\bibitem [{\citenamefont {Seidl}\ \emph {et~al.}(2017)\citenamefont {Seidl}
  \emph {et~al.}}]{Seidl:2017qhp}%
  \BibitemOpen
  \bibfield  {author} {\bibinfo {author} {\bibfnamefont {R.}~\bibnamefont
  {Seidl}} \emph {et~al.} (\bibinfo {collaboration} {Belle}),\ }\href {\doibase
  10.1103/PhysRevD.96.032005} {\bibfield  {journal} {\bibinfo  {journal} {Phys.
  Rev.}\ }\textbf {\bibinfo {volume} {D96}},\ \bibinfo {pages} {032005}
  (\bibinfo {year} {2017})},\ \Eprint {http://arxiv.org/abs/1706.08348}
  {arXiv:1706.08348 [hep-ex]} \BibitemShut {NoStop}%
\bibitem [{\citenamefont {Field}\ and\ \citenamefont
  {Feynman}(1977)}]{Field:1976ve}%
  \BibitemOpen
  \bibfield  {author} {\bibinfo {author} {\bibfnamefont {R.~D.}\ \bibnamefont
  {Field}}\ and\ \bibinfo {author} {\bibfnamefont {R.~P.}\ \bibnamefont
  {Feynman}},\ }\href {\doibase 10.1103/PhysRevD.15.2590} {\bibfield  {journal}
  {\bibinfo  {journal} {Phys. Rev.}\ }\textbf {\bibinfo {volume} {D15}},\
  \bibinfo {pages} {2590} (\bibinfo {year} {1977})}\BibitemShut {NoStop}%
\bibitem [{\citenamefont {Field}\ and\ \citenamefont
  {Feynman}(1978)}]{Field:1977fa}%
  \BibitemOpen
  \bibfield  {author} {\bibinfo {author} {\bibfnamefont {R.~D.}\ \bibnamefont
  {Field}}\ and\ \bibinfo {author} {\bibfnamefont {R.~P.}\ \bibnamefont
  {Feynman}},\ }\href {\doibase 10.1016/0550-3213(78)90015-9} {\bibfield
  {journal} {\bibinfo  {journal} {Nucl. Phys.}\ }\textbf {\bibinfo {volume}
  {B136}},\ \bibinfo {pages} {1} (\bibinfo {year} {1978})}\BibitemShut
  {NoStop}%
\bibitem [{\citenamefont {Ito}\ \emph {et~al.}(2009)\citenamefont {Ito},
  \citenamefont {Bentz}, \citenamefont {Cloet}, \citenamefont {Thomas},\ and\
  \citenamefont {Yazaki}}]{Ito:2009zc}%
  \BibitemOpen
  \bibfield  {author} {\bibinfo {author} {\bibfnamefont {T.}~\bibnamefont
  {Ito}}, \bibinfo {author} {\bibfnamefont {W.}~\bibnamefont {Bentz}}, \bibinfo
  {author} {\bibfnamefont {I.~C.}\ \bibnamefont {Cloet}}, \bibinfo {author}
  {\bibfnamefont {A.~W.}\ \bibnamefont {Thomas}}, \ and\ \bibinfo {author}
  {\bibfnamefont {K.}~\bibnamefont {Yazaki}},\ }\href {\doibase
  10.1103/PhysRevD.80.074008} {\bibfield  {journal} {\bibinfo  {journal} {Phys.
  Rev.}\ }\textbf {\bibinfo {volume} {D80}},\ \bibinfo {pages} {074008}
  (\bibinfo {year} {2009})},\ \Eprint {http://arxiv.org/abs/0906.5362}
  {arXiv:0906.5362 [nucl-th]} \BibitemShut {NoStop}%
\bibitem [{\citenamefont {Matevosyan}\ \emph
  {et~al.}(2011{\natexlab{a}})\citenamefont {Matevosyan}, \citenamefont
  {Thomas},\ and\ \citenamefont {Bentz}}]{Matevosyan:2010hh}%
  \BibitemOpen
  \bibfield  {author} {\bibinfo {author} {\bibfnamefont {H.~H.}\ \bibnamefont
  {Matevosyan}}, \bibinfo {author} {\bibfnamefont {A.~W.}\ \bibnamefont
  {Thomas}}, \ and\ \bibinfo {author} {\bibfnamefont {W.}~\bibnamefont
  {Bentz}},\ }\href {\doibase 10.1103/PhysRevD.83.074003} {\bibfield  {journal}
  {\bibinfo  {journal} {Phys.Rev.}\ }\textbf {\bibinfo {volume} {D83}},\
  \bibinfo {pages} {074003} (\bibinfo {year} {2011}{\natexlab{a}})},\ \Eprint
  {http://arxiv.org/abs/1011.1052} {arXiv:1011.1052 [hep-ph]} \BibitemShut
  {NoStop}%
\bibitem [{\citenamefont {Matevosyan}\ \emph
  {et~al.}(2011{\natexlab{b}})\citenamefont {Matevosyan}, \citenamefont
  {Thomas},\ and\ \citenamefont {Bentz}}]{Matevosyan:2011ey}%
  \BibitemOpen
  \bibfield  {author} {\bibinfo {author} {\bibfnamefont {H.~H.}\ \bibnamefont
  {Matevosyan}}, \bibinfo {author} {\bibfnamefont {A.~W.}\ \bibnamefont
  {Thomas}}, \ and\ \bibinfo {author} {\bibfnamefont {W.}~\bibnamefont
  {Bentz}},\ }\href {\doibase 10.1103/PhysRevD.83.114010,
  10.1103/PhysRevD.86.059904} {\bibfield  {journal} {\bibinfo  {journal}
  {Phys.Rev.}\ }\textbf {\bibinfo {volume} {D83}},\ \bibinfo {pages} {114010}
  (\bibinfo {year} {2011}{\natexlab{b}})},\ \Eprint
  {http://arxiv.org/abs/1103.3085} {arXiv:1103.3085 [hep-ph]} \BibitemShut
  {NoStop}%
\bibitem [{\citenamefont {Matevosyan}\ \emph
  {et~al.}(2012{\natexlab{a}})\citenamefont {Matevosyan}, \citenamefont
  {Thomas},\ and\ \citenamefont {Bentz}}]{PhysRevD.86.059904}%
  \BibitemOpen
  \bibfield  {author} {\bibinfo {author} {\bibfnamefont {H.~H.}\ \bibnamefont
  {Matevosyan}}, \bibinfo {author} {\bibfnamefont {A.~W.}\ \bibnamefont
  {Thomas}}, \ and\ \bibinfo {author} {\bibfnamefont {W.}~\bibnamefont
  {Bentz}},\ }\href {\doibase 10.1103/PhysRevD.86.059904} {\bibfield  {journal}
  {\bibinfo  {journal} {Phys. Rev. D}\ }\textbf {\bibinfo {volume} {86}},\
  \bibinfo {pages} {059904(E)} (\bibinfo {year}
  {2012}{\natexlab{a}})}\BibitemShut {NoStop}%
\bibitem [{\citenamefont {Matevosyan}\ \emph
  {et~al.}(2012{\natexlab{b}})\citenamefont {Matevosyan}, \citenamefont
  {Bentz}, \citenamefont {Cloet},\ and\ \citenamefont
  {Thomas}}]{Matevosyan:2011vj}%
  \BibitemOpen
  \bibfield  {author} {\bibinfo {author} {\bibfnamefont {H.~H.}\ \bibnamefont
  {Matevosyan}}, \bibinfo {author} {\bibfnamefont {W.}~\bibnamefont {Bentz}},
  \bibinfo {author} {\bibfnamefont {I.~C.}\ \bibnamefont {Cloet}}, \ and\
  \bibinfo {author} {\bibfnamefont {A.~W.}\ \bibnamefont {Thomas}},\ }\href
  {\doibase 10.1103/PhysRevD.85.014021} {\bibfield  {journal} {\bibinfo
  {journal} {Phys.Rev.}\ }\textbf {\bibinfo {volume} {D85}},\ \bibinfo {pages}
  {014021} (\bibinfo {year} {2012}{\natexlab{b}})},\ \Eprint
  {http://arxiv.org/abs/1111.1740} {arXiv:1111.1740 [hep-ph]} \BibitemShut
  {NoStop}%
\bibitem [{\citenamefont {Nambu}\ and\ \citenamefont
  {Jona-Lasinio}(1961{\natexlab{a}})}]{Nambu:1961tp}%
  \BibitemOpen
  \bibfield  {author} {\bibinfo {author} {\bibfnamefont {Y.}~\bibnamefont
  {Nambu}}\ and\ \bibinfo {author} {\bibfnamefont {G.}~\bibnamefont
  {Jona-Lasinio}},\ }\href {\doibase 10.1103/PhysRev.122.345} {\bibfield
  {journal} {\bibinfo  {journal} {Phys. Rev.}\ }\textbf {\bibinfo {volume}
  {122}},\ \bibinfo {pages} {345} (\bibinfo {year}
  {1961}{\natexlab{a}})}\BibitemShut {NoStop}%
\bibitem [{\citenamefont {Nambu}\ and\ \citenamefont
  {Jona-Lasinio}(1961{\natexlab{b}})}]{Nambu:1961fr}%
  \BibitemOpen
  \bibfield  {author} {\bibinfo {author} {\bibfnamefont {Y.}~\bibnamefont
  {Nambu}}\ and\ \bibinfo {author} {\bibfnamefont {G.}~\bibnamefont
  {Jona-Lasinio}},\ }\href {\doibase 10.1103/PhysRev.124.246} {\bibfield
  {journal} {\bibinfo  {journal} {Phys. Rev.}\ }\textbf {\bibinfo {volume}
  {124}},\ \bibinfo {pages} {246} (\bibinfo {year}
  {1961}{\natexlab{b}})}\BibitemShut {NoStop}%
\bibitem [{\citenamefont {Matevosyan}\ \emph {et~al.}(2013)\citenamefont
  {Matevosyan}, \citenamefont {Thomas},\ and\ \citenamefont
  {Bentz}}]{Matevosyan:2013aka}%
  \BibitemOpen
  \bibfield  {author} {\bibinfo {author} {\bibfnamefont {H.~H.}\ \bibnamefont
  {Matevosyan}}, \bibinfo {author} {\bibfnamefont {A.~W.}\ \bibnamefont
  {Thomas}}, \ and\ \bibinfo {author} {\bibfnamefont {W.}~\bibnamefont
  {Bentz}},\ }\href {\doibase 10.1103/PhysRevD.88.094022} {\bibfield  {journal}
  {\bibinfo  {journal} {Phys.Rev.}\ }\textbf {\bibinfo {volume} {D88}},\
  \bibinfo {pages} {094022} (\bibinfo {year} {2013})},\ \Eprint
  {http://arxiv.org/abs/1310.1917} {arXiv:1310.1917 [hep-ph]} \BibitemShut
  {NoStop}%
\bibitem [{\citenamefont {Matevosyan}\ \emph
  {et~al.}(2014{\natexlab{a}})\citenamefont {Matevosyan}, \citenamefont
  {Thomas},\ and\ \citenamefont {Bentz}}]{Matevosyan:2013nla}%
  \BibitemOpen
  \bibfield  {author} {\bibinfo {author} {\bibfnamefont {H.~H.}\ \bibnamefont
  {Matevosyan}}, \bibinfo {author} {\bibfnamefont {A.~W.}\ \bibnamefont
  {Thomas}}, \ and\ \bibinfo {author} {\bibfnamefont {W.}~\bibnamefont
  {Bentz}},\ }\bibfield  {booktitle} {\emph {\bibinfo {booktitle}
  {{Proceedings, 25th International Nuclear Physics Conference (INPC 2013)}}},\
  }\href {\doibase 10.1051/epjconf/20146606014} {\bibfield  {journal} {\bibinfo
   {journal} {EPJ Web Conf.}\ }\textbf {\bibinfo {volume} {66}},\ \bibinfo
  {pages} {06014} (\bibinfo {year} {2014}{\natexlab{a}})},\ \Eprint
  {http://arxiv.org/abs/1307.8125} {arXiv:1307.8125 [hep-ph]} \BibitemShut
  {NoStop}%
\bibitem [{\citenamefont {Matevosyan}\ \emph
  {et~al.}(2014{\natexlab{b}})\citenamefont {Matevosyan}, \citenamefont
  {Kotzinian},\ and\ \citenamefont {Thomas}}]{Matevosyan:2013eia}%
  \BibitemOpen
  \bibfield  {author} {\bibinfo {author} {\bibfnamefont {H.~H.}\ \bibnamefont
  {Matevosyan}}, \bibinfo {author} {\bibfnamefont {A.}~\bibnamefont
  {Kotzinian}}, \ and\ \bibinfo {author} {\bibfnamefont {A.~W.}\ \bibnamefont
  {Thomas}},\ }\href {\doibase 10.1016/j.physletb.2014.02.040} {\bibfield
  {journal} {\bibinfo  {journal} {Phys.Lett.}\ }\textbf {\bibinfo {volume}
  {B731}},\ \bibinfo {pages} {208} (\bibinfo {year} {2014}{\natexlab{b}})},\
  \Eprint {http://arxiv.org/abs/1312.4556} {arXiv:1312.4556 [hep-ph]}
  \BibitemShut {NoStop}%
\bibitem [{\citenamefont {Bentz}\ \emph {et~al.}(2016)\citenamefont {Bentz},
  \citenamefont {Kotzinian}, \citenamefont {Matevosyan}, \citenamefont
  {Ninomiya}, \citenamefont {Thomas},\ and\ \citenamefont
  {Yazaki}}]{Bentz:2016rav}%
  \BibitemOpen
  \bibfield  {author} {\bibinfo {author} {\bibfnamefont {W.}~\bibnamefont
  {Bentz}}, \bibinfo {author} {\bibfnamefont {A.}~\bibnamefont {Kotzinian}},
  \bibinfo {author} {\bibfnamefont {H.~H.}\ \bibnamefont {Matevosyan}},
  \bibinfo {author} {\bibfnamefont {Y.}~\bibnamefont {Ninomiya}}, \bibinfo
  {author} {\bibfnamefont {A.~W.}\ \bibnamefont {Thomas}}, \ and\ \bibinfo
  {author} {\bibfnamefont {K.}~\bibnamefont {Yazaki}},\ }\href {\doibase
  10.1103/PhysRevD.94.034004} {\bibfield  {journal} {\bibinfo  {journal} {Phys.
  Rev.}\ }\textbf {\bibinfo {volume} {D94}},\ \bibinfo {pages} {034004}
  (\bibinfo {year} {2016})},\ \Eprint {http://arxiv.org/abs/1603.08333}
  {arXiv:1603.08333 [nucl-th]} \BibitemShut {NoStop}%
\bibitem [{\citenamefont {Matevosyan}\ \emph
  {et~al.}(2017{\natexlab{a}})\citenamefont {Matevosyan}, \citenamefont
  {Kotzinian},\ and\ \citenamefont {Thomas}}]{Matevosyan:2016fwi}%
  \BibitemOpen
  \bibfield  {author} {\bibinfo {author} {\bibfnamefont {H.~H.}\ \bibnamefont
  {Matevosyan}}, \bibinfo {author} {\bibfnamefont {A.}~\bibnamefont
  {Kotzinian}}, \ and\ \bibinfo {author} {\bibfnamefont {A.~W.}\ \bibnamefont
  {Thomas}},\ }\href {\doibase 10.1103/PhysRevD.95.014021} {\bibfield
  {journal} {\bibinfo  {journal} {Phys. Rev.}\ }\textbf {\bibinfo {volume}
  {D95}},\ \bibinfo {pages} {014021} (\bibinfo {year} {2017}{\natexlab{a}})},\
  \Eprint {http://arxiv.org/abs/1610.05624} {arXiv:1610.05624 [hep-ph]}
  \BibitemShut {NoStop}%
\bibitem [{\citenamefont {Matevosyan}\ \emph
  {et~al.}(2017{\natexlab{b}})\citenamefont {Matevosyan}, \citenamefont
  {Kotzinian},\ and\ \citenamefont {Thomas}}]{Matevosyan:2017alv}%
  \BibitemOpen
  \bibfield  {author} {\bibinfo {author} {\bibfnamefont {H.~H.}\ \bibnamefont
  {Matevosyan}}, \bibinfo {author} {\bibfnamefont {A.}~\bibnamefont
  {Kotzinian}}, \ and\ \bibinfo {author} {\bibfnamefont {A.~W.}\ \bibnamefont
  {Thomas}},\ }\href {\doibase 10.1103/PhysRevD.96.074010} {\bibfield
  {journal} {\bibinfo  {journal} {Phys. Rev.}\ }\textbf {\bibinfo {volume}
  {D96}},\ \bibinfo {pages} {074010} (\bibinfo {year} {2017}{\natexlab{b}})},\
  \Eprint {http://arxiv.org/abs/1707.04999} {arXiv:1707.04999 [hep-ph]}
  \BibitemShut {NoStop}%
\bibitem [{\citenamefont {Boer}\ \emph {et~al.}(2003)\citenamefont {Boer},
  \citenamefont {Jakob},\ and\ \citenamefont {Radici}}]{Boer:2003ya}%
  \BibitemOpen
  \bibfield  {author} {\bibinfo {author} {\bibfnamefont {D.}~\bibnamefont
  {Boer}}, \bibinfo {author} {\bibfnamefont {R.}~\bibnamefont {Jakob}}, \ and\
  \bibinfo {author} {\bibfnamefont {M.}~\bibnamefont {Radici}},\ }\href
  {\doibase 10.1103/PhysRevD.67.094003} {\bibfield  {journal} {\bibinfo
  {journal} {Phys.Rev.}\ }\textbf {\bibinfo {volume} {D67}},\ \bibinfo {pages}
  {094003} (\bibinfo {year} {2003})},\ \Eprint
  {http://arxiv.org/abs/hep-ph/0302232} {arXiv:hep-ph/0302232 [hep-ph]}
  \BibitemShut {NoStop}%
\bibitem [{\citenamefont {Gliske}\ \emph {et~al.}(2014)\citenamefont {Gliske},
  \citenamefont {Bacchetta},\ and\ \citenamefont {Radici}}]{Gliske:2014wba}%
  \BibitemOpen
  \bibfield  {author} {\bibinfo {author} {\bibfnamefont {S.}~\bibnamefont
  {Gliske}}, \bibinfo {author} {\bibfnamefont {A.}~\bibnamefont {Bacchetta}}, \
  and\ \bibinfo {author} {\bibfnamefont {M.}~\bibnamefont {Radici}},\ }\href
  {\doibase 10.1103/PhysRevD.90.114027, 10.1103/PhysRevD.91.019902} {\bibfield
  {journal} {\bibinfo  {journal} {Phys. Rev.}\ }\textbf {\bibinfo {volume}
  {D90}},\ \bibinfo {pages} {114027} (\bibinfo {year} {2014})},\ \bibinfo
  {note} {[Erratum: Phys. Rev.D91,no.1,019902(2015)]},\ \Eprint
  {http://arxiv.org/abs/1408.5721} {arXiv:1408.5721 [hep-ph]} \BibitemShut
  {NoStop}%
\bibitem [{\citenamefont {Ceccopieri}\ \emph {et~al.}(2007)\citenamefont
  {Ceccopieri}, \citenamefont {Radici},\ and\ \citenamefont
  {Bacchetta}}]{Ceccopieri:2007ip}%
  \BibitemOpen
  \bibfield  {author} {\bibinfo {author} {\bibfnamefont {F.~A.}\ \bibnamefont
  {Ceccopieri}}, \bibinfo {author} {\bibfnamefont {M.}~\bibnamefont {Radici}},
  \ and\ \bibinfo {author} {\bibfnamefont {A.}~\bibnamefont {Bacchetta}},\
  }\href {\doibase 10.1016/j.physletb.2007.04.065} {\bibfield  {journal}
  {\bibinfo  {journal} {Phys.Lett.}\ }\textbf {\bibinfo {volume} {B650}},\
  \bibinfo {pages} {81} (\bibinfo {year} {2007})},\ \Eprint
  {http://arxiv.org/abs/hep-ph/0703265} {arXiv:hep-ph/0703265 [HEP-PH]}
  \BibitemShut {NoStop}%
\bibitem [{\citenamefont {Abdesselam}\ \emph {et~al.}(2015)\citenamefont
  {Abdesselam} \emph {et~al.}}]{Abdesselam:2015nxn}%
  \BibitemOpen
  \bibfield  {author} {\bibinfo {author} {\bibfnamefont {A.}~\bibnamefont
  {Abdesselam}} \emph {et~al.} (\bibinfo {collaboration} {Belle}),\ }\href@noop
  {} {\  (\bibinfo {year} {2015})},\ \Eprint {http://arxiv.org/abs/1505.08020}
  {arXiv:1505.08020 [hep-ex]} \BibitemShut {NoStop}%
\bibitem [{\citenamefont {Vossen}(2015)}]{Vossen:2015znm}%
  \BibitemOpen
  \bibfield  {author} {\bibinfo {author} {\bibfnamefont {A.}~\bibnamefont
  {Vossen}},\ }\bibfield  {booktitle} {\emph {\bibinfo {booktitle}
  {{Proceedings, 23rd International Workshop on Deep-Inelastic Scattering and
  Related Subjects (DIS 2015): Dallas, Texas, USA, April 27-May 01, 2015}}},\
  }\href@noop {} {\bibfield  {journal} {\bibinfo  {journal} {PoS}\ }\textbf
  {\bibinfo {volume} {DIS2015}},\ \bibinfo {pages} {216} (\bibinfo {year}
  {2015})}\BibitemShut {NoStop}%
\bibitem [{\citenamefont {Sirtl}(2017)}]{Sirtl:2017rhi}%
  \BibitemOpen
  \bibfield  {author} {\bibinfo {author} {\bibfnamefont {S.}~\bibnamefont
  {Sirtl}},\ }in\ \href
  {http://inspirehep.net/record/1514894/files/arXiv:1702.07317.pdf} {\emph
  {\bibinfo {booktitle} {{22nd International Symposium on Spin Physics (SPIN
  2016) Urbana, IL, USA, September 25-30, 2016}}}}\ (\bibinfo {year} {2017})\
  \Eprint {http://arxiv.org/abs/1702.07317} {arXiv:1702.07317 [hep-ex]}
  \BibitemShut {NoStop}%
\bibitem [{\citenamefont {Bacchetta}\ and\ \citenamefont
  {Radici}(2006)}]{Bacchetta:2006un}%
  \BibitemOpen
  \bibfield  {author} {\bibinfo {author} {\bibfnamefont {A.}~\bibnamefont
  {Bacchetta}}\ and\ \bibinfo {author} {\bibfnamefont {M.}~\bibnamefont
  {Radici}},\ }\href {\doibase 10.1103/PhysRevD.74.114007} {\bibfield
  {journal} {\bibinfo  {journal} {Phys.Rev.}\ }\textbf {\bibinfo {volume}
  {D74}},\ \bibinfo {pages} {114007} (\bibinfo {year} {2006})},\ \Eprint
  {http://arxiv.org/abs/hep-ph/0608037} {arXiv:hep-ph/0608037 [hep-ph]}
  \BibitemShut {NoStop}%
\end{thebibliography}%

\end{document}